\documentclass{elsarticle}

\usepackage[margin=2cm]{geometry}
\usepackage{graphicx}
\usepackage{lmodern}
\usepackage{amsmath}
\usepackage{amsfonts}
\usepackage{bm}
\usepackage{hyperref}
\usepackage{mathtools}
\usepackage{booktabs}

\hypersetup{unicode=true,colorlinks=true,linkcolor=blue,urlcolor=blue,citecolor=blue,breaklinks=true}

\renewcommand{\vec}[1]{{\bm{#1}}}
\newcommand{\tens}[1]{\mathbb{#1}}
\newcommand{\CBar}{\overline{\tens{C}}}
\newcommand{\kBar}{\overline{\bm{k}}}
\newcommand{\alphaBar}{\overline{\bm{\alpha}}}
\newcommand{\eps}{\bm{\varepsilon}}
\newcommand{\sig}{\bm{\sigma}}
\newcommand{\epsBar}{\overline{\bm{\varepsilon}}}
\newcommand{\sigBar}{\overline{\bm{\sigma}}}
\newcommand{\vf}{\mathrm{vf}}
\newcommand{\ar}{\mathrm{ar}}
\newcommand{\vfop}{\operatorname{\mathsf{vf}}}
\newcommand{\DMN}{\operatorname{\mathrm{DMN}}}
\newcommand{\Lam}{\operatorname{\mathrm{Lam}}}
\newcommand{\vtheta}{\vec{\theta}}
\newcommand{\DMNparams}{(\vec{w},\vec{\theta})}
\newcommand{\nnodes}{N}
\newcommand{\card}{\operatorname{\mathrm{card}}}
\newcommand{\saij}{\vec{a}}
\newcommand{\saijop}{\vec{\mathsf{a}}}
\newcommand{\valpha}{\vec{\alpha}}
\newcommand{\DeltaT}{\Delta T}

\let\oldeqref\eqref
\renewcommand{\eqref}[1]{Eq.~\oldeqref{#1}}

\DeclarePairedDelimiter{\abs}{\lvert}{\rvert}
\DeclarePairedDelimiter{\norm}{\lVert}{\rVert}

\begin{document}
\title{Micromechanics-Informed Parametric Deep Material Network for Physics Behavior Prediction of Heterogeneous Materials with a Varying Morphology}

\author[DS]{Tianyi Li}
\ead{tianyi.li@3ds.com}
\address[DS]{Dassault Systèmes, 10 rue Marcel Dassault, 78140 Vélizy-Villacoublay, France}

\begin{abstract}
    Deep Material Network (DMN) has recently emerged as a data-driven surrogate model for heterogeneous materials. Given a particular microstructural morphology, the effective linear and nonlinear behaviors can be successfully approximated by such physics-based neural-network like architecture. In this work, a novel micromechanics-informed parametric DMN (MIpDMN) architecture is proposed for multiscale materials with a varying microstructure characterized by several parameters. A single-layer feedforward neural network is used to account for the dependence of DMN fitting parameters on the microstructural ones. Micromechanical constraints are prescribed both on the architecture and the outputs of this new neural network. The proposed MIpDMN is also recast in a multiple physics setting, where physical properties other than the mechanical ones can also be predicted. In the numerical simulations conducted on three parameterized microstructures, MIpDMN demonstrates satisfying generalization capabilities when morphology varies. The effective behaviors of such parametric multiscale materials can thus be predicted and encoded by MIpDMN with high accuracy and efficiency.
\end{abstract}


\maketitle

\section{Introduction}
The effective macroscopic behaviors of heterogeneous materials can be predicted by the physical properties of the constituents and the underlying microstructures using a multiscale modeling approach \cite{Kanoute:2009,Matous:2017,Fish:2021}. Among such models, mean-field homogenization schemes rely on the micromechanical assumptions (inclusion shapes, interaction between the constituents) of the microstructure being considered and are frequently used to predict linear and nonlinear behaviors of specific composites (such as the fiber-reinforced plastics) \cite{TuckerIII:1999,Pierard:2004}. Thanks to their analytical nature, they are computationally efficient and can be used as a local constitutive model on each integration point for concurrent structural-scale simulation of industrial components. However, their inherent simplifying hypotheses on the idealized microstructure may not be appropriate for other complex morphologies. In this case, full-field simulation of the micromechanical problem would be necessary, following a computational homogenization approach \cite{Geers:2010}. For instance, finite element (FE) simulation of a Representative Volume Element (RVE) of the underlying microstructure can be carried out to compute the local solution fields. The accurate linear and nonlinear effective responses can then be obtained through field averaging. However, such full-field method involves much higher computational cost and is impractical to be run concurrently in multiscale simulations on industrial parts.

Recently, the Deep Material Network (DMN) method \cite{Liu:2019} has emerged as a novel data-driven surrogate model for such multiscale materials. Based on linear elastic training data generated by high-fidelity computational homogenization (FE-RVE simulation, for instance), DMN is capable of predicting their elastic and inelastic behaviors with high accuracy and efficiency even in the finite-strain range. Compared to other data-driven approaches for material modeling \cite{Kumar:2022} that rely on the (effective) strain-stress pair data, DMN learns instead the morphological characterizations of the microstructure via its homogenization function $(\tens{C}_1,\tens{C}_2)\mapsto\CBar$. Here, $(\tens{C}_1,\tens{C}_2)$ are the stiffness tensors of the constituents and $\CBar$ is the effective stiffness tensor. Due to its physics-based neural-network like architecture, DMN demonstrates great expressive power with much fewer fitting parameters compared to traditional machine learning methods. Thanks to these interesting properties, it is now gaining popularity in the computational mechanics community:
\begin{itemize}
    \item In \cite{Liu:2019a}, DMN has been successfully applied for 3-d particle-reinforced composites, polycrystalline materials and woven composites involving three modeling scales. Nonlinear complex material behaviors such as hyperelasticity with stress softening and rate-dependent crystal plasticity are investigated under a finite-strain setting.
    \item A new rotation-free formulation along with a flattened architecture is proposed in \cite{Gajek:2020}. It has been tested for short-fiber plastics and metal matrix composites with an elasto-plastic behavior in the matrix phase.
    \item Applications to short-fiber reinforced materials are also considered further in \cite{Gajek:2021,Gajek:2022,Wei:2023}. The thermomechanical behaviors of such composites can be successfully captured by DMN. Concurrent multiscale simulations combining a macroscopic finite element model and DMN on each integration point are performed for static and dynamic problems on structural components.
    \item More complex material behaviors such as viscoplasticity or failure are investigated in \cite{Liu:2020,Liu:2021,Dey:2022,Dey:2023}. New training strategies are also proposed to obtain more accurate nonlinear responses.
    \item In \cite{Wu:2021}, other DMN architectures are explored with mean-field homogenization based building blocks. Applications to woven microstructures are considered. A generalization of the original DMN framework based on interactions between discrete material nodes is also proposed in \cite{Nguyen:2022,Nguyen:2022a}. It has been successfully applied to porous microstructures with an elasto-plastic behavior.
\end{itemize}

Despite these successes indicated by numerical evidences, the expressive power of DMN remains to be fully understood from a theoretical perspective. In \cite{Gajek:2020}, its micromechanical justifications are provided in the context of generalized standard materials \cite{Halphen:1975}. Thanks to the hierarchical architecture based on an appropriate laminate microstructure as the building block, DMN is thermodynamically consistent and verifies the micromechanical bounds on the effective linear and nonlinear behaviors. Actually, DMN can be identified as a multiple-rank laminate microstructure in homogenization theory \cite{Allaire:2002,Milton:2002}, in which each constituent of a laminate is also a laminate. Sequential laminates (also called coated laminates) are a subset among such hierarchical microstructures, where one of the constituents in the lamination process is always the same. It is well known that the Hashin-Shtrikman bounds \cite{Hashin:1962} can be attained by such finite-rank sequential laminates. In the case of an effectively isotropic composite of two isotropic phases, such bounds can be realized by a rank-3 laminate in 2-d or a rank-6 laminate for 3-d elasticity \cite{Francfort:1986}. As noted in \cite[pg. 294]{Milton:2002}, sequential laminates can approximate the homogenization function $(\tens{C}_1,\tens{C}_2)\mapsto\CBar$ associated with a two-phase microstructure to the second order in $\tens{C}_1-\tens{C}_2$. Since sequential laminates are a proper subset of multi-rank laminates, these results can be naturally extended to DMN and hence provide some preliminary theoretical justifications of its expressive power.

In general, we are often dealing with a class of microstructures that present similar morphologies. These microstructures are characterized by several parameters defining the geometrical shapes of each constituent inside the RVE. For instance, the volume fraction parameter $\vf$ describes the overall relative volume of each constituent. Other parameters also exist, depending on the exact morphological characterization of the microstructure. These micromechanical parameters may have an important influence on the effective properties. Hence, it would be beneficial that DMN also captures such structure-property relationships with high accuracy and efficiency. In \cite{Liu:2019b,Huang:2022}, a transfer-learning based approach is proposed to construct a DMN database for such parameterized microstructures. Training of several DMN instances is performed sequentially following a pre-determined path in the parametric space. Piecewise linear interpolation is then used to include the dependence of DMN fitting parameters on the microstructural parameters. In their work, they only focus on microstructures that can be characterized by $\vf$ and a second-order orientation tensor \cite{Advani:1987}.

Meanwhile, a regression-based approach is proposed in \cite{Gajek:2021} for short-fiber reinforced materials with a constant fiber volume fraction. The fitting parameters of one single DMN instance are assumed dependent on the varying (principal) fiber orientation tensor $(a_1,a_2,a_3)$. The offline training is performed using a total loss function that aggregates multiple microstructures with different $(a_1,a_2,a_3)$ values. The main objective of this paper is to extend this formulation to generic parameterized microstructures, with varying volume fractions and possibly other geometrical parameters. In particular, we investigate of the use of a single-layer feedforward neural network to account for the dependence of DMN parameters on the microstructural ones. Additional micromechanics-based constraints are included to improve generalization capability of our parametric DMN architecture.

Another objective of this paper is to recast DMN in a multiple physics setting. In the literature, DMN has been used extensively as a surrogate of the possibly nonlinear \emph{mechanical} behaviors of heterogeneous materials. Other physical behaviors have not been explored except in \cite{Gajek:2022}, where a two-way coupled thermomechanical problem is considered. The temperature-dependent stress-strain behaviors as well as the mechanically-induced self-heating are predicted. However, effective physical properties, such as thermal conductivity or the coefficient of thermal expansion, are not explicitly computed by DMN. In this paper, we demonstrate that physical properties other than the mechanical ones can also be accurately predicted by DMN, even though it is trained using isothermal linear elastic mechanical data. It should be noted that multiphysics coupling is not considered in this paper.

The paper is organized as follows. The original DMN formulation \cite{Liu:2019,Liu:2019a} is first reviewed and then recast in a multiple physics setting in Sect. \ref{sec:dmn}. Extensions to parameterized microstructures are then considered in Sect. \ref{sec:parametric}, in which we introduce a new micromechanics-informed parametric DMN architecture. The proposed method is further evaluated on three parameterized microstructures in Sect. \ref{sec:numerical}. The obtained numerical results highlight the effects of the introduced physical constraints. Finally, the conclusions from the current work are summarized in Sect. \ref{sec:conclusion}.

\section{Deep Material Network} \label{sec:dmn}
\begin{figure}[htbp]
    \centering
    \includegraphics[width=0.5\textwidth]{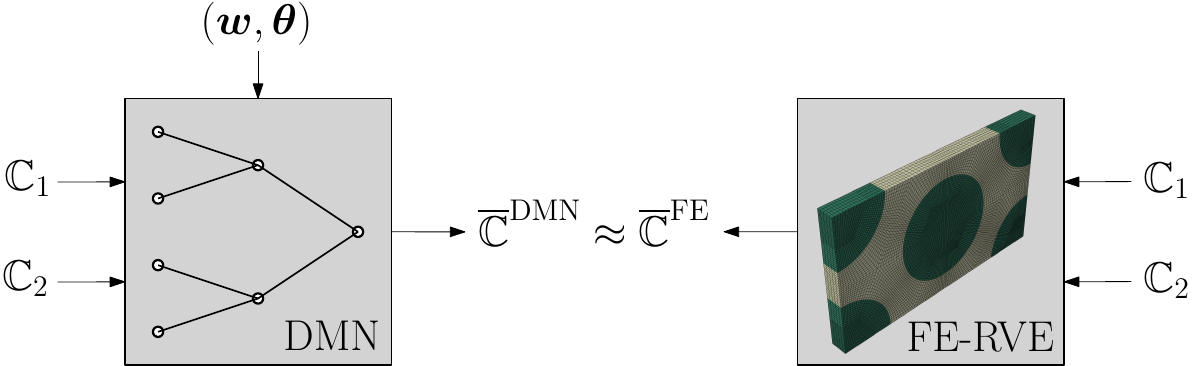}
    \caption{DMN learns the morphology of a particular microstructure $\Omega$ through its homogenization function.} \label{fig:dmn_ferve}
\end{figure}

Instead of learning the macroscopic (nonlinear) strain-stress behaviors of a particular two-phase microstructure $\Omega$, as shown in Fig. \ref{fig:dmn_ferve}, DMN learns its linear elastic homogenization function
\begin{equation} \label{eq:dmn_C1C2CBar}
    (\tens{C}_1,\tens{C}_2)\mapsto\CBar.
\end{equation}
The effective computation of \eqref{eq:dmn_C1C2CBar} on complex microstructures $\Omega$ is in general realized by computational homogenization like FE-RVE methods. The homogenization function encodes in particular the microstructure morphology, which can be defined by the characteristic function
\begin{equation} \label{eq:chi}
    \chi(\vec{x})=\begin{cases}
        1 & \vec{x}\in\text{Phase 1}, \\
        0 & \text{otherwise}.
    \end{cases}
\end{equation}
In this regard, DMN is not a surrogate of the \emph{effective behaviors} of the microstructure but the microstructure \emph{per se}.

\subsection{Network architecture and fitting parameters} \label{sec:network}
In this work, we follow the original formulation of \cite{Liu:2019,Liu:2019a} based on a perfect binary tree architecture and a rotation-based formulation for laminates, with minor modifications that will be explained in the sequel.

\begin{figure}[htbp]
    \centering
    \includegraphics[width=0.9\textwidth]{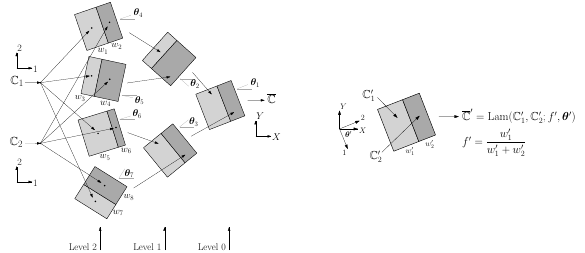}
    \caption{An $L$-layer DMN architecture can be represented by a perfect binary tree with each ``node'' being a rank-1 laminate microstructure. The DMN material nodes $1\leq i\leq\nnodes=2^L$ are represented by dots $\cdot$ on the input layer and receive the linear elastic properties $(\tens{C}_1,\tens{C}_2)$ of the two phases. In this example with 3 DMN layers, three nesting levels is involved in order to compute the homogenization function $\DMN_3:(\tens{C}_1,\tens{C}_2)\mapsto\CBar$. The fitting parameters are the weights $\vec{w}$ defined on the leaf laminates and the rotations $\vtheta$ on each laminate.} \label{fig:dmn_laminate}
\end{figure}

DMN is a multiple-rank laminate microstructure as defined in \cite[Chap. 9]{Milton:2002}, comprised of hierarchically nested laminates of laminates on different length scales, see Fig. \ref{fig:dmn_laminate}. The rank $L$ of such multiple-rank laminates, also called the number of DMN layers, characterizes the number of nesting levels. Its architecture corresponds topologically to a perfect binary tree. Each ``node'' is a rank-1 laminate microstructure, serving as the ``mechanistic building blocks'' \cite{Liu:2019,Liu:2019a,Gajek:2020} or ``neurons'' in this neural-network like architecture.

On the input layer, each leaf laminate of the binary tree receives the linear elastic properties of the two phases $(\tens{C}_1,\tens{C}_2)$ expressed in their respective material frames. All these $\nnodes=2^L$ entries are called DMN material nodes, since each of them will be associated with a constitutive model of the phase 1 or 2 for online multiple physics behavior prediction. The index set of the material nodes will be denoted $\mathbb{I}$. It can be partitioned evenly into two subsets for each of the phase.
\begin{equation}
    \mathbb{I}=\{1,\ldots,\nnodes\}=\mathbb{I}_1\cup\mathbb{I}_2,\quad\mathbb{I}_1=\{2j-1\mid j=1,\ldots,2^{L-1}\},\quad\mathbb{I}_2=\{2j\mid j=1,\ldots,2^{L-1}\}.
\end{equation}

On the output layer, the effective properties of the microstructure $\CBar$ are predicted through the root laminate and are given in the global frame of the microstructure $\Omega$. The homogenization function $\DMN_L:(\tens{C}_1,\tens{C}_2)\mapsto\CBar$ of such $L$-layer DMN can be defined recursively as follows
\begin{equation} \label{eq:DMN_recursive}
    \CBar=\DMN_L(\tens{C}_1,\tens{C}_2)=\begin{cases}
        \Lam(\tens{C}_1,\tens{C}_2)                                                         & L=1, \\
        \Lam\bigl(\DMN_{L-1}(\tens{C}_1,\tens{C}_2),\DMN_{L-1}(\tens{C}_1,\tens{C}_2)\bigr) & L>1.
    \end{cases}
\end{equation}
In \eqref{eq:DMN_recursive}, each $\Lam$ denotes the homogenization function of the rank-1 laminate associated with different weights $(w_1',w_2')$ and rotations $\vtheta'$, see Fig. \ref{fig:dmn_laminate}. Compared to the original formulation \cite{Liu:2019}, we omit the material rotation matrices for phase constitutive properties on the input layer. Note that by omitting these input rotations, the material frames of the input properties will actually coincide with the local laminate frames.

A rank-1 laminate is a microstructure in which its characteristic function \eqref{eq:chi} only varies in the direction of lamination $\chi(\vec{x})=\chi(\vec{x}\cdot\vec{n})$. In 3-d, its homogenization function $\Lam:(\tens{C}_1',\tens{C}_2')\mapsto\CBar'$ can thus be characterized by only two parameters: the volume fraction $f'$ of one of the phases, and a $3\times 3$ rotation matrix $\vec{R}(\vtheta')\in\mathrm{SO}(3)$, where $\mathrm{SO}(3)$ is the special orthogonal group acting on $\mathbb{R}^3$. The primes on the parameters and on the input and output tensors indicate that they don't represent necessarily the properties of the constituents (except for the input layer), nor the effective behaviors of the microstructure (except for the output layer). The constituent properties $(\tens{C}_1',\tens{C}_2')$ are expressed in the laminate frame $(\vec{e}_1,\vec{e}_2,\vec{e}_3)$, see Fig. \ref{fig:dmn_laminate}. The lamination direction $\vec{n}$ aligns with this local frame: $\vec{e}_2$ for 2-d microstructures \cite{Liu:2019}, and $\vec{e}_3$ for 3-d ones \cite{Liu:2019a}. The homogenized behaviors of the laminate are first computed locally, then rotated to the frame of another laminate on the next nesting level or the global frame for the laminate on level 0. Thanks to the simplifying geometry, the homogenization function $\Lam$ is analytical.

In summary, the fitting parameters of an $L$-layer DMN are
\begin{itemize}
    \item Weights $\vec{w}=(w_1,\ldots,w_\nnodes)$, defined on the leaf laminates. Its length $\nnodes=2^L$ is the number of DMN material nodes. They are required to be non-negative, hence an adequate nonlinear activation function $\sigma$ is applied
          \begin{equation} \label{eq:w_from_z}
              \vec{w}=\sigma(\vec{z})\geq 0,\quad \vec{z}\in\mathbb{R}^\nnodes.
          \end{equation}
          ReLU $\sigma(z)=\max(0,z)$ is used in \cite{Liu:2019,Liu:2019a,Gajek:2020,Gajek:2021,Wu:2021}, while Softplus $\sigma(z)=\log(1+\exp(\beta z))/\beta$ is used in \cite{Nguyen:2022,Nguyen:2022a} as a smooth approximation to ReLU.

          These weights are propagated from the leaf laminates to the root so that each laminate receives different $(w_1',w_2')$, see Fig. \ref{fig:dmn_laminate}. Denoting the weights on level $0\leq i< L$ by $\vec{w}'^{(i)}$, with $\vec{w}'^{(L-1)}=\vec{w}$ the weights on the leaf laminates, we have
          \begin{equation} \label{eq:weight_propagation}
              \vec{w}'^{(i)}_j=\vec{w}'^{(i+1)}_{2j-1}+\vec{w}'^{(i+1)}_{2j},\quad j=1,\ldots,2^{i+1}.
          \end{equation}
          For each laminate, the volume fraction $f'$ can then be computed
          \begin{equation} \label{eq:volume_fraction}
              f'=\frac{w_1'}{w_1'+w_2'}.
          \end{equation}
          Due to \eqref{eq:weight_propagation} and \eqref{eq:volume_fraction}, the DMN homogenization function \eqref{eq:DMN_recursive} is invariant with respect to a scaling on DMN weights $\vec{w}\mapsto k\vec{w}$ for $k>0$.

    \item Rotations $\vtheta=(\vtheta_1,\ldots,\vtheta_{2^L-1})$, defined for each laminate. Its length $2^L-1$ equals to the number of ``nodes'' in a perfect binary tree. Its elements can be organized using breadth-first ordering like in Fig. \ref{fig:dmn_laminate}. Each element $\vtheta_i$ represents a $3\times 3$ rotation matrix $\vec{R}(\vtheta_i)$. In \cite{Liu:2019a,Wu:2021}, the three Euler angles are used. In this work however, we use the quaternion representation which avoids singularity issues with Euler angles \cite{Diebel:2006}. Hence, $\vtheta$ is a $(2^L-1)\times 4$ matrix.
\end{itemize}
The total number of fitting parameters can be found in Tab. \ref{tab:num_dmn_parameters}.
\begin{table}[htbp]
    \centering
    \begin{tabular}{llll} \toprule
        $L$-layer DMN                & Weights       & Rotations           & Total           \\ \midrule
        Number of fitting parameters & $\nnodes=2^L$ & $4\times (2^{L}-1)$ & $5\times 2^L-4$ \\ \bottomrule
    \end{tabular}
    \caption{Number of fitting parameters for an $L$-layer DMN, using the network architecture explained in Sect. \ref{sec:network} with in particular the quaternion representation for the rotation matrices.}
    \label{tab:num_dmn_parameters}
\end{table}

The training of these fitting parameters follows a data-driven approach, by comparing the DMN prediction of the homogenized stiffness tensor and that predicted by computational homogenization, see Fig. \ref{fig:dmn_ferve}. The machine learning procedure is described in \cite{Liu:2019a,Gajek:2020} and will be presented in Sect. \ref{sec:training} for our parametric DMN architecture.

After training, DMN not only provides an accurate and efficient surrogate of the linear homogenization function \eqref{eq:DMN_recursive}, it serves also as a surrogate microstructure to predict its nonlinear behaviors. In \ref{sec:nonlinear}, the nonlinear online prediction architecture of DMN \cite{Liu:2019} is recalled. In addition, we introduce and numerically investigate the use of acceleration methods \cite{Ramiere:2015} to further improve computational efficiency.

\subsection{Multiple physics property prediction} \label{sec:multiphi}
As noted at the beginning of Sect. \ref{sec:dmn} and in Fig. \ref{fig:dmn_ferve}, DMN should not be regarded as a surrogate of the \emph{effective mechanical behaviors} but the microstructure \emph{per se}. It is hence desirable that other physical properties can be predicted along with the mechanical ones at the online prediction stage. In this paper, we propose to consider the following two additional physical properties
\begin{enumerate}
    \item Thermal conductivity $\vec{k}$, which is a symmetric second-order tensor for anisotropic behaviors.
    \item Coefficient of thermal expansion $\bm{\alpha}$, also a symmetric second-order tensor in the anisotropic case.
\end{enumerate}

It turns out that the online prediction of $\vec{k}$ and $\bm{\alpha}$ requires only an adequate redefinition of the building block (neuron), denoted by $\Lam$ in \eqref{eq:DMN_recursive}. In particular, the DMN network architecture as well as the fitted parameters after offline training remain exactly the same. This property naturally recasts DMN in a multiple physics setting. Although each DMN neuron corresponds to a laminate microstructure, the exact formula of $\Lam$ will now depend on the physics being considered. Below, we will provide the definition of $\Lam$ for these additional physics properties. For completeness, its original definition \cite{Liu:2019,Liu:2019a} for (linear elastic) mechanical behaviors is also included.

To ease the notation, we will now ignore the primes on the input and output tensors in the laminate homogenization function, see Fig. \ref{fig:dmn_ferve}. For instance, for linear elasticity we now have
\[
    \CBar=\Lam_\tens{C}(\tens{C}_1,\tens{C}_2).
\]
It should be remembered that the input tensors don't necessarily refer to those of the two constituents (except for the input layer), and the output effective tensor is not necessarily the final effective property tensor of the microstructure (except for the output layer).

\subsubsection*{Linear elastic behaviors}
The linear elastic behaviors of the laminate microstructure is governed by static equilibrium. Based on the lamination direction $\vec{n}$, the stress components can be partitioned into two parts: a tangential part $\sig^\mathrm{t}$ and a normal part $\sig^\mathrm{n}$. For example, if the lamination direction aligns with the local $\vec{e}_3$ vector, then in Mandel notation we have
\begin{equation} \label{eq:laminationtn}
    \sig^\mathrm{t}=(\sigma_{11},\sigma_{22},\sqrt{2}\sigma_{12}),\quad\sig^\mathrm{n}=(\sigma_{33},\sqrt{2}\sigma_{13},\sqrt{2}\sigma_{23}).
\end{equation}
The strain tensor as well as the stiffness tensors can be partitioned similarly. The interface condition prescribes that the normal part of the stress tensor $\sig^\mathrm{n}$ is continuous, while the tangential part of the strain tensor $\eps^\mathrm{t}$ is continuous. It can be encoded by the following matrix equation
\begin{equation} \label{eq:interface}
    \begin{bmatrix}
        \eps_1^\mathrm{t} \\ \sig_1^\mathrm{n}
    \end{bmatrix}=
    \begin{bmatrix}
        \eps_2^\mathrm{t} \\ \sig_2^\mathrm{n}
    \end{bmatrix}
    \implies
    \underbrace{\begin{bmatrix}
            \mathbf{I}_{3\times 3}     & \vec{0}_{3\times 3}        \\
            \mathbb{C}_1^{\mathrm{nt}} & \mathbb{C}_1^{\mathrm{nn}}
        \end{bmatrix}}_{\widehat{\mathbb{C}}_1}
    \begin{bmatrix}
        \eps_1^\mathrm{t} \\ \eps_1^\mathrm{n}
    \end{bmatrix}=
    \underbrace{\begin{bmatrix}
            \mathbf{I}_{3\times 3}     & \vec{0}_{3\times 3}        \\
            \mathbb{C}_2^{\mathrm{nt}} & \mathbb{C}_2^{\mathrm{nn}}
        \end{bmatrix}}_{\widehat{\mathbb{C}}_2}
    \begin{bmatrix}
        \eps_2^\mathrm{t} \\ \eps_2^\mathrm{n}
    \end{bmatrix},
\end{equation}
where $\mathbb{C}_1$ and $\mathbb{C}_2$ are the $6\times 6$ stiffness tensors of the two laminate phases in Mandel notation.

Using the definition of the effective strain tensor $\epsBar=f\eps_1+(1-f)\eps_2$, where $f$ is the volume fraction of the phase 1, we obtain the following formula that computes $\eps_1$ from $\epsBar$
\begin{equation}
    \bigl((1-f)\widehat{\mathbb{C}}_1+f\widehat{\mathbb{C}}_2\bigr)\eps_1=\widehat{\mathbb{C}}_2\epsBar\implies \eps_1=\mathbb{A}\epsBar=\widehat{\mathbb{C}}^{-1}\widehat{\mathbb{C}}_2\epsBar,\quad \widehat{\mathbb{C}}=(1-f)\widehat{\mathbb{C}}_1+f\widehat{\mathbb{C}}_2,
\end{equation}
where $\mathbb{A}=\widehat{\mathbb{C}}^{-1}\widehat{\mathbb{C}}_2$ is the strain localization tensor.

Finally, with the effective stress tensor $\sigBar=f\sig_1+(1-f)\sig_2=f\mathbb{C}_1\eps_1+(1-f)\mathbb{C}_2\eps_2$, the effective stiffness tensor can be obtained as follows
\begin{equation} \label{eq:laminateCBar}
    \sigBar=\CBar\epsBar,\quad \CBar=f(\mathbb{C}_1-\mathbb{C}_2)\mathbb{A}+\mathbb{C}_2.
\end{equation}
The rotation can then be applied adequately for the fourth-order stiffness tensor $\CBar$.

\subsubsection*{Thermal conductivity}
Now, the neuron homogenization function computes the second-order effective (anisotropic) conductivity tensor $\kBar$ from those of its constituents $\vec{k}_1$ and $\vec{k}_2$
\[
    \kBar=\Lam_{\vec{k}}(\vec{k}_1,\vec{k}_2).
\]
The laminate microstructure is governed by steady-state heat equation and the Fourier's law $\vec{q}=-\vec{k}\nabla T$ that relates the temperature gradient $\nabla T$ to the heat flux $\vec{q}$ through the conductivity tensor. As in \eqref{eq:laminationtn}, the heat flux and the temperature gradient can be decomposed into a tangential and a normal part
\[
    \vec{q}^\mathrm{t}=(q_1,q_2),\quad \vec{q}^\mathrm{n}=(q_3).
\]
Similar to \eqref{eq:interface}, the interface condition can be described by the following linear system
\[
    \begin{bmatrix}
        \nabla T_1^\mathrm{t} \\ \vec{q}_1^\mathrm{n}
    \end{bmatrix}=
    \begin{bmatrix}
        \nabla T_2^\mathrm{t} \\ \vec{q}_2^\mathrm{n}
    \end{bmatrix}
    \implies
    \underbrace{\begin{bmatrix}
            \mathbf{I}_{2\times 2}  & \vec{0}_{2\times 1}     \\
            \vec{k}_1^{\mathrm{nt}} & \vec{k}_1^{\mathrm{nn}}
        \end{bmatrix}}_{\widehat{\vec{k}}_1}
    \begin{bmatrix}
        \nabla T_1^\mathrm{t} \\ \nabla T_1^\mathrm{n}
    \end{bmatrix}=
    \underbrace{\begin{bmatrix}
            \mathbf{I}_{2\times 2}  & \vec{0}_{2\times 1}     \\
            \vec{k}_2^{\mathrm{nt}} & \vec{k}_2^{\mathrm{nn}}
        \end{bmatrix}}_{\widehat{\vec{k}}_2}
    \begin{bmatrix}
        \nabla T_2^\mathrm{t} \\ \nabla T_2^\mathrm{n}
    \end{bmatrix}.
\]

Following the same procedure as before, the effective conductivity tensor is given by
\begin{equation} \label{eq:laminatekBar}
    \overline{\vec{q}}=-\kBar\,\overline{\nabla T},\quad \kBar=f(\vec{k}_1-\vec{k}_2)\mathbf{A}+\vec{k}_2,\quad \mathbf{A}=\widehat{\vec{k}}^{-1}\widehat{\vec{k}}_2,\quad \widehat{\vec{k}}=(1-f)\widehat{\vec{k}}_1+f\widehat{\vec{k}}_2.
\end{equation}
The rotation can be applied to the second-order tensor $\kBar$.

\subsubsection*{Coefficient of thermal expansion}
The effective coefficient of thermal expansion (CTE), as a second-order tensor $\alphaBar$ in the anisotropic case, can be computed along with the effective stiffness tensor. Now, the laminate homogenization function becomes
\[
    (\CBar,\alphaBar)=\Lam_{\tens{C},\valpha}(\tens{C}_1,\tens{C}_2,\valpha_1,\valpha_2).
\]
While $\CBar$ is still given by \eqref{eq:laminateCBar}, the effective CTE can be obtained by using the relationship \cite{Rosen:1970,Benveniste:1990} between $\alphaBar$ and the effective compliance tensor $\overline{\tens{S}}=\CBar^{-1}$. For completeness, its derivation is now recalled below.

Consider an arbitrary two-phase microstructure with fixed material orientation. It is subjected to a uniform temperature difference $\DeltaT$ and a traction vector $\sigBar\vec{n}$ on its boundary, such that the strain, stress and temperature fields are \emph{uniform} inside the microstructure. Using the thermoelastic constitutive equation, we obtain
\begin{equation} \label{eq:thermoelastic1}
    \eps_1=\eps_2=\tens{S}_1\sigBar+\valpha_1\DeltaT=\tens{S}_2\sigBar+\valpha_2\DeltaT\implies \sigBar=(\tens{S}_1-\tens{S}_2)^{-1}(\valpha_2-\valpha_1)\DeltaT.
\end{equation}
In \eqref{eq:thermoelastic1}, similar to stiffness and compliance tensors, the second-order CTE tensors are also expressed in Mandel notation by a $6\times 1$ vector. The expression $\tens{S}^{-1}\valpha=\tens{C}\valpha$ is hence understood as matrix multiplication similar to $\tens{C}\eps$.

The volume-averaged stress $\sigBar_1$ of the phase 1 can be related to $\sigBar$ and $\DeltaT$ by using the mechanical stress concentration tensor $\tens{B}$ and the thermal stress concentration tensor $\vec{b}$
\begin{equation} \label{eq:thermoelastic_phase1}
    \sigBar_1=\tens{B}\sigBar+\vec{b}\DeltaT.
\end{equation}
Under the previous thermoelastic loading, using \eqref{eq:thermoelastic1}, we obtain hence
\begin{equation} \label{eq:bfromB}
    \vec{b}=(\vec{I}_{6\times 6}-\tens{B})(\tens{S}_1-\tens{S}_2)^{-1}(\valpha_2-\valpha_1).
\end{equation}
This universal relationship \eqref{eq:bfromB} is valid for arbitrary thermoelastic loading of the microstructure. It implies that the thermal stress concentration tensor $\vec{b}$ can be uniquely determined as long as the mechanical one is given.

The effective thermoelastic constitutive equation for the microstructure is given by
\begin{equation} \label{eq:thermoelastic_constitutive}
    \epsBar=\overline{\tens{S}}\sigBar+\alphaBar\DeltaT.
\end{equation}
With \eqref{eq:thermoelastic_phase1} under isothermal loading $\DeltaT=0$, the effective compliance tensor can be computed by
\begin{equation} \label{eq:SBarfromB}
    \overline{\tens{S}}=f(\tens{S}_1-\tens{S}_2)\tens{B}+\tens{S}_2,
\end{equation}
where $f$ is the volume fraction of the phase 1 as before. Combining \eqref{eq:thermoelastic_phase1} to \eqref{eq:SBarfromB} with $\sigBar=f\sig_1+(1-f)\sig_2=\vec{0}$, we obtain
\begin{equation} \label{eq:epsBaralphaBar}
    \begin{aligned}
        \epsBar=\alphaBar\DeltaT & = f\eps_1+(1-f)\eps_2=\bigl(f\valpha_1+(1-f)\valpha_2\bigr)\DeltaT+f(\tens{S}_1-\tens{S}_2)\sig_1                                                                   \\
                                 & = \bigl(f\valpha_1+(1-f)\valpha_2\bigr)\DeltaT+\bigl(\overline{\tens{S}}-f\tens{S}_1-(1-f)\tens{S}_2\bigr)(\tens{S}_1-\tens{S}_2)^{-1}(\valpha_1-\valpha_2)\DeltaT.
    \end{aligned}
\end{equation}
After some manipulations, we deduce that the effective CTE tensor reads
\begin{equation} \label{eq:laminatealphaBar}
    \alphaBar=\valpha_1+(\overline{\tens{S}}-\tens{S}_1)(\tens{S}_1-\tens{S}_2)^{-1}(\valpha_1-\valpha_2).
\end{equation}

Note that \eqref{eq:laminatealphaBar} is not only valid for each laminate microstructure (DMN neuron) but also for the microstructure being approximated by DMN, as long as material orientation is fixed for both phases. In the latter case, first $\CBar$ is predicted by using \eqref{eq:DMN_recursive}, then \eqref{eq:laminatealphaBar} can be directly applied using $\overline{\tens{S}}=\CBar^{-1}$ computed by DMN. The only requirement is that the (effective) thermoelastic constituent properties are expressed in the same global frame.

When at least one of the phases contains varying local material orientation, \eqref{eq:laminatealphaBar} needs to be applied as a \emph{neuron} operation similar to \eqref{eq:laminateCBar} and \eqref{eq:laminatekBar}, from the input layer to the output layer. For each laminate microstructure, the following operations are expressed and performed in the same laminate frame:
\begin{enumerate}
    \item Compute $\CBar$ using \eqref{eq:laminateCBar}.
    \item Compute $\alphaBar$ using \eqref{eq:laminatealphaBar} and $\overline{\tens{S}}=\CBar^{-1}$ just obtained.
    \item Apply appropriate rotation operations for $\CBar$ and $\alphaBar$.
\end{enumerate}
Rotations must be applied at the end, since \eqref{eq:laminatealphaBar} is only valid when expressed in the same frame. The rotated $\CBar$ and $\alphaBar$ then become the inputs for a laminate of the next nesting level (or directly the final DMN outputs).

\section{Extension to parameterized microstructures} \label{sec:parametric}
\begin{figure}[htbp]
    \centering
    \includegraphics[width=0.6\textwidth]{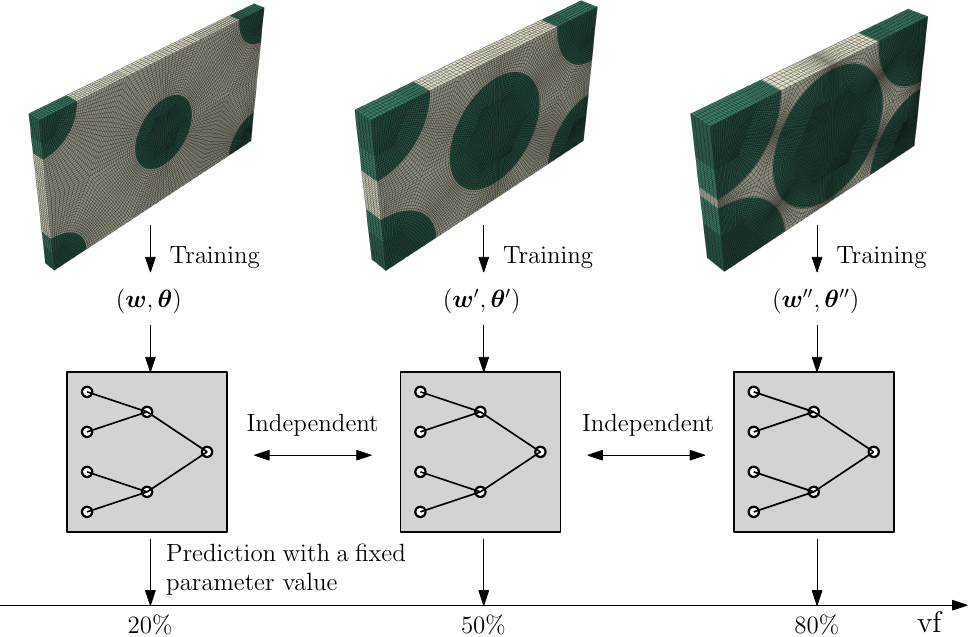}
    \caption{With the original formulation, each DMN is trained independently to learn a particular microstructure $\Omega_{\vec{p}_i}$.} \label{fig:dmn_multiple_parameters}
\end{figure}

In general, we are often dealing with a class of morphologically similar microstructures, described by one or several parameters $\vec{p}=(p_1,p_2,\ldots)$. In the case of the unidirectional fiber composite, illustrated in Fig. \ref{fig:dmn_multiple_parameters}, they are parameterized by the single $\vf$ parameter which indicates the volume fraction of the fibers.

The original DMN formulation \cite{Liu:2019} is not adapted for such parameterized microstructures $\vec{p}\mapsto\Omega_\vec{p}$, since each DMN is trained to learn a particular microstructure $\Omega_{\vec{p}_i}$ with a fixed parameter value $\vec{p}_i$. After independent training on such $n$ microstructures, we only obtain a \emph{discrete} set of DMN objects
\[
    \{\DMN_{(\vec{w}_i,\vec{\theta}_i)},\quad i=1,\ldots, n\}.
\]
Each $\DMN_{(\vec{w}_i,\vec{\theta}_i)}$ is tailored to approximate the homogenized behavior of a \emph{particular} microstructure $\Omega_{\vec{p}_i}$. Without another training (which requires the costly computational homogenization data), it is not possible to predict the behavior of a new microstructure $\Omega_{\vec{p}'}$ that is absent in the previous set.

\subsection{Transfer-learning based interpolative DMN} \label{sec:transfer_learning}
\begin{figure}[htbp]
    \centering
    \includegraphics[width=0.6\textwidth]{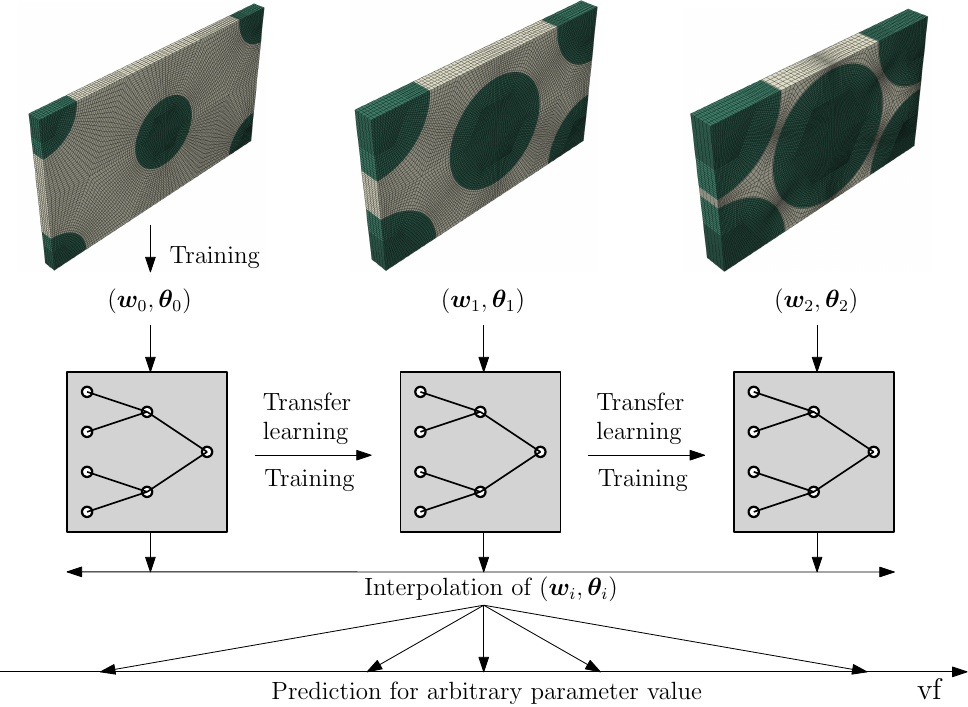}
    \caption{Transfer-learning based interpolative deep material network.} \label{fig:dmn_transfer_learning}
\end{figure}

In \cite{Liu:2019b}, a transfer-learning strategy is proposed to interpolate different DMN models trained at different parameter values $\vec{p}_i$. The functional dependence $\vec{p}\mapsto\bigl(\vec{w}(\vec{p}),\vec{\theta}(\vec{p})\bigr)$ is defined by interpolating the DMN parameters $(\vec{w}_i,\vec{\theta}_i)$ between different microstructural parameters $\vec{p}_i$. This assumes implicitly that the same DMN architecture with the same number of layers is used. If each DMN is trained independently using random initialization of its fitting parameters, $\vec{p}\mapsto\bigl(\vec{w}(\vec{p}),\vec{\theta}(\vec{p})\bigr)$ would not be \emph{smooth}, which leads to less accurate predictions when interpolating between known $\vec{p}_i$. This motivates hence a transfer-learning based training strategy, illustrated in Fig. \ref{fig:dmn_transfer_learning}.

A \emph{pre-determined sequential} path $(\vec{p}_0,\vec{p}_1,\ldots,\vec{p}_n)$ in the parametric space is required to carry out transfer-learning training on DMN parameters
\begin{equation*}
    \DMN_{(\vec{w}_0,\vec{\theta}_0)}\to\DMN_{(\vec{w}_1,\vec{\theta}_1)}\to\cdots\to\DMN_{(\vec{w}_n,\vec{\theta}_n)}.
\end{equation*}
The training of $(\vec{w}_i,\vec{\theta}_i)$ is initialized using previously trained $(\vec{w}_{i-1},\vec{\theta}_{i-1})$, for $i>0$. Only the first training at $\vec{p}_0$ is performed using standard random initialization. Transfer learning not only ensures a smooth interpolation between DMN parameters, it also accelerates training of subsequent DMN models for new microstructural parameters $\vec{p}_i$.

While this transfer-learning based approach has been tested with success for microstructures depending solely on the volume fraction parameter \cite{Liu:2019b}, and those depending on the volume fraction and the orientation tensor \cite{Huang:2022}. In our opinion it has also some limitations:
\begin{itemize}
    \item Definition of a \emph{pre-determined training sequence} may become difficult especially for higher parametric dimensions. For microstructures with multiple geometrical parameters of different natures, it is not trivial to choose the starting point $\vec{p}_0$ as well as the sequential interpolation points $\vec{p}_i$. Furthermore, the expressive power of DMN in the parametric space may also depend on the chosen training path $(\vec{p}_0,\vec{p}_1,\ldots,\vec{p}_n)$. The number of DMN active material nodes (those with a positive weight) may decrease in the process of transfer learning with the ReLU activation function. An example is given in \cite{Liu:2019b}.
    \item Extrapolation especially for higher parametric dimensions. Based on interpolation points, it would be difficult to extrapolate ``outside'' the training domain. In \cite{Liu:2019b,Huang:2022} for instance, extrapolation in the parametric space is not considered.
    \item Computational efficiency concerns. Since transfer learning is sequential in nature, DMN at different microstructural parameters can only be trained one after another. Also, each DMN instance associated with its parameters needs to be stored for interpolation.
\end{itemize}

\subsection{Micromechanics-informed parametric DMN} \label{sec:pinndmn}
In this work, we propose a novel micromechanics-informed parametric DMN (MIpDMN) architecture dedicated for parameterized microstructures. The functional dependence of DMN parameters $(\vec{w},\vtheta)$ on the microstructural parameters $\vec{p}$ is now directly accounted for by a single-layer feedforward neural network. We obtain thus
\begin{subequations} \label{eq:fully_connected}
    \begin{align}
        \vec{w}(\vec{p}) & = \sigma(\vec{W}_1\vec{p}+\vec{w}_0), \label{eq:wWpz}   \\
        \vtheta(\vec{p}) & = \vec{\Theta}_1\vec{p}+\vec{\theta}_0, \label{eq:tTpt}
    \end{align}
\end{subequations}
where $\vec{W}_1$ and $\vec{\Theta}_1$ are "weight" tensors of appropriate dimensions, $\vec{w}_0$ and $\vec{\theta}_0$ are ``bias'' tensors and $\sigma$ denotes a nonlinear activation function as in \eqref{eq:w_from_z}. Several remarks are in order:
\begin{itemize}
    \item The tensors $\vec{W}_1$ and $\vec{\Theta}_1$ characterize the dependence of DMN parameters $\vec{w}$ and $\vtheta$ on microstructural parameters, while $\vec{w}_0$ and $\vec{\theta}_0$ remain constant while microstructural parameters change.
    \item When the microstructural parameters have little effect on the effective behaviors of the microstructures, we would have $\vec{W}_1=\vec{\Theta}_1=\vec{0}$ and thus recover the original DMN formulation \cite{Liu:2019}. The ``bias'' parameters $\vec{w}_0$ and $\vec{\theta}_0$ solely are needed to approximate the homogenized behavior of this parameterized microstructure.
    \item Due to the presence of a nonlinear activation function applied after affine transformation, the functional dependence of the DMN weights \eqref{eq:wWpz} is different from piecewise linear interpolation used in the transfer-learning based interpolative DMN approach \cite{Liu:2019b,Huang:2022}. Even when the ReLU activation function is used, \eqref{eq:wWpz} implies that each $w_i$ is \emph{individually} piecewise linear with different slope-changing points. In \cite{Liu:2019b,Huang:2022} however, linear interpolation is performed \emph{globally} on the DMN weights vector $\vec{w}$.
    \item Similar functional dependence for the DMN rotations \eqref{eq:tTpt} has been proposed in \cite{Gajek:2021}. They have shown that an affine dependence works best in terms of accuracy and generalization capabilities, in comparison with other nonlinear functions. Hence, hidden layers as well as nonlinear activation functions are not considered for $\vtheta(\vec{p})$ in this work.
\end{itemize}

\begin{figure}[htbp]
    \centering
    \includegraphics[width=0.5\textwidth]{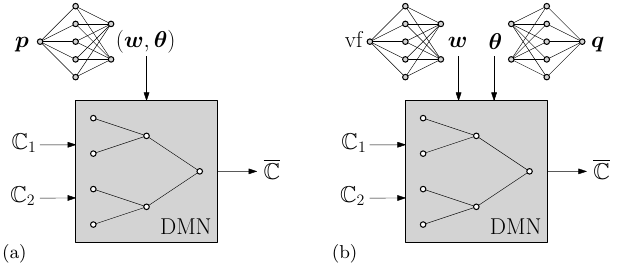}
    \caption{Parametric deep material network: (a) fully-connected architecture; (b) micromechanics-informed one.} \label{fig:dmn_scheme_parametric}
\end{figure}

The neural network proposed in \eqref{eq:fully_connected} follows a \emph{fully-connected} architecture in the sense that the DMN weights $\vec{w}$ and rotations $\vec{\theta}$ depend on \emph{all} the microstructural parameters. Meanwhile, several findings in the literature motivate a \emph{micromechanics-informed} (MI) architecture by separating the dependence of DMN weights and rotations:
\begin{itemize}
    \item For parameterized microstructures with a single \emph{volume fraction} parameter, the naïve approach, as proposed by \cite{Liu:2019b} in which only the DMN weights vary with $\vf$ and the DMN rotations remain constant, actually predicts physically plausible $\vf$-dependence of the homogenized behaviors.
    \item For microstructures with parameters that do not change the volume fraction, such as for short-fiber reinforced composites with a fixed volume fraction but different fiber orientations, \cite{Gajek:2021} indicates that only the DMN rotations need to vary with such parameters, with a constant DMN weights vector.
\end{itemize}
It should be noted that these results originate directly from the micromechanics-based design of DMN \cite{Gajek:2020}. In order that the parametric DMN reflects these micromechanical properties, the microstructural parameters are partitioned
\begin{equation} \label{eq:pvq}
    \vec{p}=(\vf,\vec{q})\in\mathbb{R}\times\mathbb{R}^q,
\end{equation}
where $\vf$ is the volume fraction of the phase 2, while $\vec{q}$ denotes all other $q$ \emph{independent} purely morphological parameters that are \emph{orthogonal to} $\vf$. Given the partition \eqref{eq:pvq}, the MI architecture is now given by
\begin{subequations} \label{eq:physics_informed}
    \begin{align}
        \vec{w}(\vec{p}) & = \vec{w}(\vf)=\sigma(\vf\cdot\vec{w}_1+\vec{w}_0), \label{eq:pwWpz}      \\
        \vtheta(\vec{p}) & = \vtheta(\vec{q})=\vec{\Theta}_1\vec{q}+\vec{\theta}_0, \label{eq:ptTpt}
    \end{align}
\end{subequations}
where $\vec{w}_0$ and $\vec{w}_1$ are vectors of length $\nnodes$, $\vec{\theta}_0$ is a $(2^L-1)\times 4$ matrix and $\vec{\Theta}_1$ is a $(2^L-1)\times 4\times q$ tensor. The fully-connected architecture \eqref{eq:fully_connected} and the MI one \eqref{eq:physics_informed} are compared in Fig. \ref{fig:dmn_scheme_parametric}.

The partition \eqref{eq:pvq} can be made more mathematically precise through the $n$-point correlation functions \cite{Gokhale:2005} which provide a hierarchy of statistical descriptions of the microstructure. The $\vf$ parameter only quantifies the relative volume proportion of each phase and corresponds to the 1-point correlation function. The other $q$ purely morphological parameters do not change the current volume fraction (which is specified by $\vf$) and contain information from higher-order correlation functions (2-point, 3-point, etc.). The \emph{orthogonal} partition can be achieved through a principal component analysis \cite{Niezgoda:2011,Pathan:2019,Sengodan:2021} on such correlation functions. The $\vf$ parameter can be extracted as a major principal feature orthogonal to other principal parameters $\vec{q}$ that are purely morphological.

The advantages of the MI architecture \eqref{eq:physics_informed} compared to the fully-connected one is two-fold. Firstly, fewer fitting parameters are required due to the partition \eqref{eq:pvq}, which increases computational efficiency. Secondly, as we shall see through numerical examples, it provides comparable expressive power and may also enhance generalization ability in the parametric space.

\begin{figure}[htbp]
    \centering
    \includegraphics[width=0.6\textwidth]{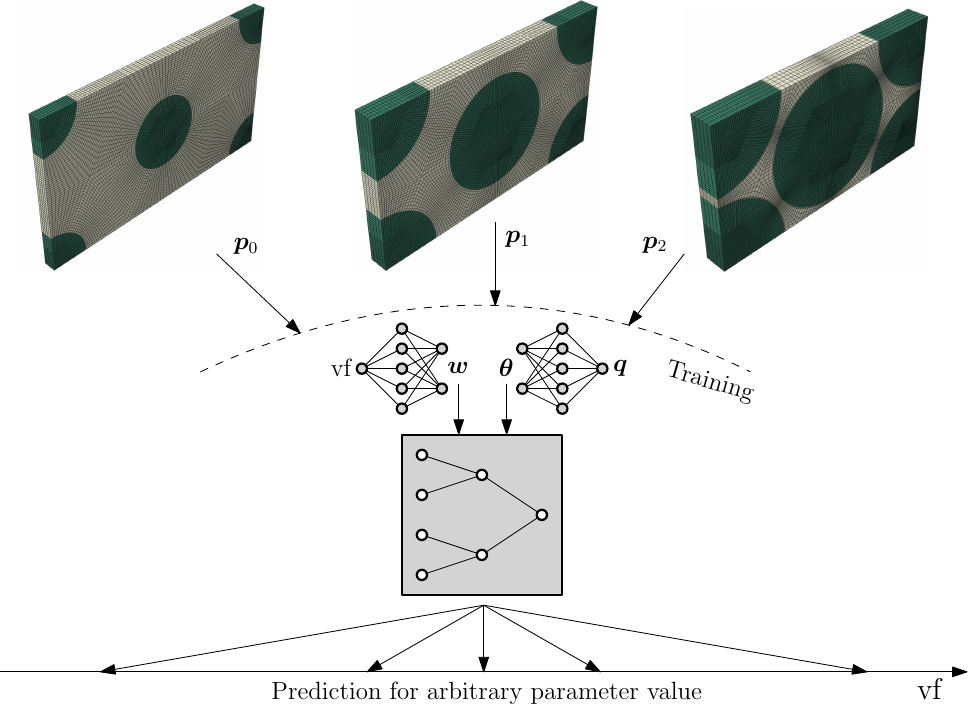}
    \caption{Offline training for MIpDMN based on the linear elastic behavior data at each $\vec{p}_i$ in the parametric space.} \label{fig:dmn_parametric}
\end{figure}

Compared to the transfer learning-based interpolative DMN described in Sect. \ref{sec:transfer_learning}, a \emph{unique} offline training is now required to optimize the fitting parameters of MIpDMN, see Fig. \ref{fig:dmn_parametric}. The expressive power of DMN is evaluated \emph{jointly} using the linear elastic behavior data at each $\vec{p}_i$ in the parametric space. Furthermore, a neural-network functional dependence naturally defines interpolation and extrapolation inside or outside the training domain and extends easily to higher parametric dimensions.

This micromechanics-informed parametric DMN will be denoted in the sequel by MIpDMN. To summarize, it is defined by:
\begin{itemize}
    \item DMN homogenization function \eqref{eq:DMN_recursive} which learns the morphology of a particular microstructure $\Omega_\vec{p}$.
    \item Micromechanics-informed single-layer feedforward neural network \eqref{eq:physics_informed} which learns the functional dependence of DMN parameters on microstructural parameters $\vec{p}$. It contains the fitting parameters of MIpDMN: $(\vec{w}_0,\vec{w}_1,\vtheta_0,\vec{\Theta}_1)$.
\end{itemize}

With an $L$-layer architecture, the number of fitting parameters can be found in Tab. \ref{tab:num_dmn_parameters_pinn}. The offline training of these fitting parameters follows a data-driven machine learning approach which will be described in Sect. \ref{sec:training}.
\begin{table}[htbp]
    \centering
    \begin{tabular}{lll} \toprule
        $L$-layer MIpDMN             & Weights            & Rotations                \\ \midrule
        Number of fitting parameters & $2\nnodes=2^{L+1}$ & $4(q+1)\times (2^{L}-1)$ \\ \bottomrule
    \end{tabular}
    \caption{Fitting parameters and their numbers for an $L$-layer MIpDMN.}
    \label{tab:num_dmn_parameters_pinn}
\end{table}

\subsection{Volume fraction constraint} \label{sec:vf_constraint}
Apart from the MI architecture on the functional dependence $\vec{p}\mapsto\DMNparams$, some physical constraints can also be prescribed on the outputs of this neural network to further improve its generalization ability, similar to the physics-informed machine learning approach for nonlinear partial differential equations \cite{Raissi:2019}.

It is now well known that the DMN weights vector $\vec{w}$ reflects the actual volume fraction information of the microstructure $\Omega$ being considered \cite{Liu:2019,Gajek:2020}. Specifically, the volume fraction of the phase 2 learned by DMN is given by
\begin{equation} \label{eq:dmn_vf}
    \vfop(\vec{w})=\frac{\sum_{i\in\mathbb{I}_2}w_i}{\sum_{i\in\mathbb{I}} w_i}\approx \vf_\Omega.
\end{equation}
In \cite{Dey:2022}, this physical property is explicitly included in the loss function as an additional constraint in learning a particular microstructure. Using our MI functional dependence \eqref{eq:pwWpz}, \eqref{eq:dmn_vf} can be further enforced at \emph{all volume fraction} values
\begin{equation} \label{eq:pinndmn_vf}
    \vfop(\vec{w})=\vfop\bigl(\sigma(\vf\cdot\vec{w}_1+\vec{w}_0)\bigr)=\vf,\quad\forall \vf\in[0,1].
\end{equation}
This volume fraction constraint is prescribed on DMN weights $\vec{w}$.

Due to the presence of a nonlinear activation function $\sigma$, \eqref{eq:pinndmn_vf} is nonlinear and the unknowns $(\vec{w}_0,\vec{w}_1)$ can not be solved explicitly. Furthermore, and luckily enough since it contributes to the expressive power of MIpDMN, there may exist multiple solutions that satisfy \eqref{eq:pinndmn_vf}. Anticipating the fact that this constraint will also be included during the offline training of our MIpDMN, we adopt a machine-learning approach to weakly enforce this volume fraction constraint through a loss function
\begin{equation} \label{eq:pinndmn_vf_loss}
    \mathcal{L}_{\vf}=\frac{1}{n_{\mathrm{v}}}\sum_{i=1}^{n_{\mathrm{v}}}\left(\vfop\bigl(\sigma(\vf_i\cdot\vec{w}_1+\vec{w}_0)\bigr)-\vf_i\right)^2,
\end{equation}
where $\vf_i$ denotes the collocation points at which \eqref{eq:pinndmn_vf} is weakly prescribed and $n_{\mathrm{v}}$ is the number of such collocation points. The sampling of these collocation points, the initialization of $(\vec{w}_0,\vec{w}_1)$ and the optimization algorithm will be described together with the offline training of MIpDMN in Sect. \ref{sec:training}.

\begin{figure}[htbp]
    \centering
    \includegraphics[width=0.8\textwidth]{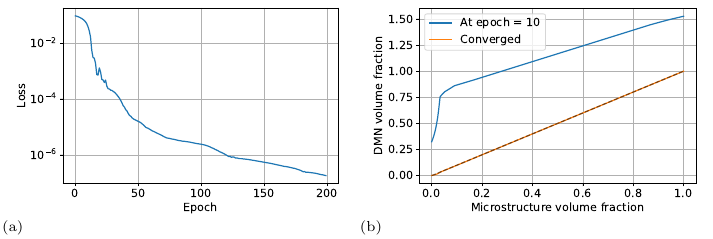}
    \caption{Volume fraction constraint prescribed on $\vec{w}(\vec{p})$: (a) minimization of the loss function \eqref{eq:pinndmn_vf_loss}; (b) predicted DMN volume fraction at the early stage and at the end of training.} \label{fig:nn_weights_training}
\end{figure}

Using 5 DMN layers and the ReLU activation function, an example of the minimization of \eqref{eq:pinndmn_vf_loss} is shown in Fig. \ref{fig:nn_weights_training}. The loss function is decreasing, which demonstrates that our MI functional dependence \eqref{eq:pwWpz} of $\vec{w}$ is capable of satisfying the volume fraction constraint \eqref{eq:pinndmn_vf}. In this example, after about 200 epochs, the maximum absolute error of $\abs{\vfop(\vec{w})-\vf}$ in $[0, 1]$ becomes approximately 0.5\%.

Another physical information that can be extracted from \eqref{eq:pwWpz} is the number of active DMN material nodes. For arbitrary $\vf\in[0,1]$, the number of active nodes for the phase $p\in\{1,2\}$ is given by
\begin{equation*}
    \card\{i\in\mathbb{I}_p\mid w_i(\vf)>0\},
\end{equation*}
where $\card(\cdot)=\abs{\cdot}$ is the cardinality of a set. In practice, machine epsilon is used instead of 0. The ratio of active DMN nodes for each phase can be obtained by dividing their respective number of active nodes by $\frac{1}{2}\nnodes$, which is the number of total material nodes per phase. In the whole parametric space, the number of \emph{globally} active nodes is defined by
\begin{equation*}
    \card\{i\in\mathbb{I}\mid w_i(\vf)>0,\quad\forall\vf\in[0,1]\}.
\end{equation*}

\begin{figure}[htbp]
    \centering
    \includegraphics[width=0.4\textwidth]{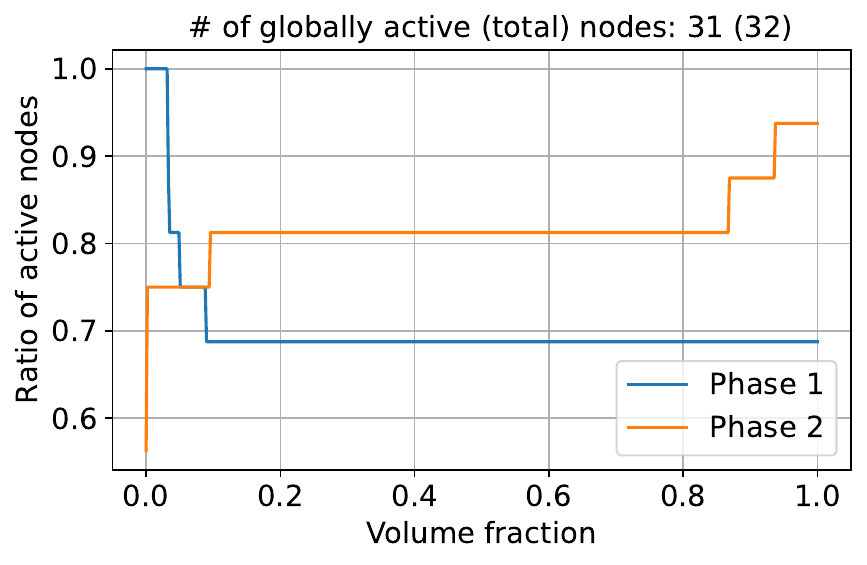}
    \caption{Variation of the ratio of active DMN nodes with the volume fraction of the phase 2.} \label{fig:nn_weights_nodes_active}
\end{figure}

In Fig. \ref{fig:nn_weights_nodes_active}, these quantities are plotted as a function of $\vf$, the volume fraction of the phase 2. In order to satisfy \eqref{eq:pinndmn_vf}, the ratio of active nodes of the phase 1 decreases from 100\% to less than 70\%, while for the phase 2 this ratio increases from less than 60\% up to more than 90\%. Our MI functional dependence \eqref{eq:pwWpz} of $\vec{w}$ is hence capable of adapting the weights of each phase as a function of the microstructure volume fraction.

\subsection{Orientation constraint} \label{sec:aij_constraint}
DMN not only captures the microstructure morphology, it also learns material orientation distribution function $\Omega\to\mathrm{SO}(3)$ in the microstructure \cite{Liu:2019a}. Using orientation tensors, we propose an orientation constraint to be prescribed on DMN weights $\vec{w}$ and rotations $\vtheta$, in order that the parametric DMN generalizes such material orientation knowledge in the whole parametric space.

Orientation tensors \cite{Advani:1987} describe concisely the statistical information of the orientation distribution of \emph{unit vectors}. Given an orientation distribution function $f:\mathbb{S}^2\to\mathbb{R}$, where $\mathbb{S}^2$ denotes the two-dimensional surface of a unit sphere, the $3\times 3$ second-order orientation tensor is defined by
\begin{equation} \label{eq:aij}
    \vec{a}=\int_{\mathbb{S}^2}f(\vec{e})\vec{e}\otimes\vec{e}\,\mathrm{d}S,\quad \norm{\vec{e}}=1,\quad (\vec{e}\otimes\vec{e})_{ij}=e_ie_j.
\end{equation}
It can be easily shown that $\vec{a}$ is symmetric and $\operatorname{tr}(\vec{a})=1$, due to the normalization constraints of $\vec{e}$ and of the probability density $f$. Higher order orientation tensors do exist \cite{Bauer:2021}, however the second-order one \eqref{eq:aij} is the most frequently used to characterize local fiber orientations due to manufacturing processes and their influence on material properties \cite{Mueller:2016,Bauer:2023}.

\begin{figure}[htbp]
    \centering
    \includegraphics[width=0.98\textwidth]{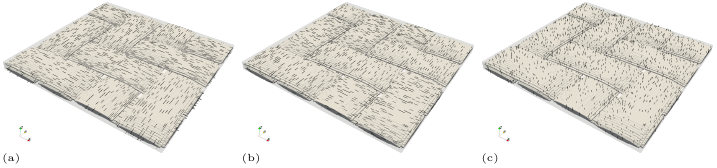}
    \caption{Orientation distribution of the material frame of the tows in the woven microstructure: (a) $\vec{e}_1$; (b) $\vec{e}_2$; (c) $\vec{e}_3$.} \label{fig:woven_e1e2e3}
\end{figure}

In this work, we propose a generalization of such (second-order) orientation tensors for orientation distributions of \emph{rotations}. Contrary to transversely isotropic fibers for which a single unit vector suffices to characterize its material frame, general anisotropic materials (like orthotropic ones) require a rotation matrix $\vec{R}$ to describe the transformation from its material frame $(\vec{e}_1,\vec{e}_2,\vec{e}_3)$ to the global one. In such cases, the orientation distribution function is now defined for rotation matrices $f:\mathrm{SO}(3)\to\mathbb{R}$. Such probability density function can also be found for the texture analysis of polycrystalline materials \cite{Boehlke:2005} or rigid body dynamics \cite{Lee:2008}. Given that each column $\vec{R}^{(i)}$ of $\vec{R}$ essentially expresses $\vec{e}_i\in\mathbb{S}^2$ in the global frame, it defines an orientation distribution of the material frame, cf. Fig. \ref{fig:woven_e1e2e3} for the tows in this woven microstructure.

Given this interpretation, three (second-order) orientation tensors can be defined for each axis of the material frame
\begin{equation} \label{eq:aij_3}
    \saij^{(i)}=\int_{\mathrm{SO}(3)}f(\vec{R})\vec{R}^{(i)}\otimes\vec{R}^{(i)}\,\mathrm{d}R,
\end{equation}
where $\mathrm{d}R$ is the volume element for $\mathrm{SO}(3)$ and Einstein summation is not implied. For discrete probability functions defined on a mesh, the integral can be understood as weighted-averaging using the element volumes as weights. Note that in the case of a two-phase microstructure, \eqref{eq:aij_3} can be computed for each of the phase. Each of these orientation tensors satisfies $\operatorname{tr}(\saij^{(i)})=1$. Due to the orthonormality of $(\vec{e}_1,\vec{e}_2,\vec{e}_3)$, we also have
\begin{equation*}
    \saij^{(1)}+\saij^{(2)}+\saij^{(3)}=\begin{bmatrix}
        1 &   &   \\
          & 1 &   \\
          &   & 1
    \end{bmatrix}.
\end{equation*}
Using \eqref{eq:aij_3} in the case of Fig. \ref{fig:woven_e1e2e3}, the material frame orientation tensors for the tows are given by
\begin{equation} \label{eq:saij_tow}
    \saij^{(1)}=\begin{bmatrix}
        0.5 & 0   & 0 \\
        0   & 0.5 & 0 \\
        0   & 0   & 0
    \end{bmatrix},\quad \saij^{(2)}=\begin{bmatrix}
        0.5 & 0   & 0 \\
        0   & 0.5 & 0 \\
        0   & 0   & 0
    \end{bmatrix},\quad \saij^{(3)}=\begin{bmatrix}
        0 & 0 & 0 \\
        0 & 0 & 0 \\
        0 & 0 & 1
    \end{bmatrix}.
\end{equation}
The tows are thus isotropically oriented in the $X$-$Y$ plane for $\vec{e}_1$ and $\vec{e}_2$, and unidirectionally oriented in the $Z$ axis for $\vec{e}_3$. For the matrix phase in Fig. \ref{fig:woven_e1e2e3}, its material frame coincides with the global frame. In this case, $\saij$ is similar to the unidirectional orientation tensor and satisfies
\begin{equation} \label{eq:saij_ud}
    a^{(i)}_{ii}=1,\quad a^{(i)}_{jk}=0\quad\text{for other components}.
\end{equation}

Suggested by \cite{Huang:2022}, such orientation tensors can also be defined for DMN. As explained in Sect. \ref{sec:network}, DMN rotations $\vtheta$ define the rotation matrices between the local frame of the current laminate to that of another laminate on the next nesting level. They can thus be composed to obtain the effective rotation matrix from the material frame $(\vec{e}_1,\vec{e}_2,\vec{e}_3)$ (leaf laminates) to the global frame of the microstructure $\Omega$ (root laminate). Let $\mathsf{p}(i)$ denote the parent of a laminate $i$ in the DMN binary tree architecture. For instance, in the 3-layer DMN example shown in Fig. \ref{fig:dmn_laminate}, we have $\mathsf{p}(4)=2$ for the leaf laminate 4 and $\mathsf{p}^2(4)=\mathsf{p}\bigl(\mathsf{p}(4)\bigr)=\mathsf{p}(2)=1$ which is the root laminate. For each leaf laminate $i$ which carries the DMN material nodes, the effective rotation matrix from the material frame to the global one is given by
\begin{equation}
    \tilde{\vec{R}}_i=\vec{R}(\vtheta_{\mathsf{p}^{L-1}(i)})\cdots\vec{R}(\vtheta_{\mathsf{p}^2(i)})\vec{R}(\vtheta_{\mathsf{p}(i)})\vec{R}(\vtheta_i),\quad \tilde{\vec{R}}_i\in\mathrm{SO}(3).
\end{equation}
Note that for an $L$-layer DMN, we have necessarily $\mathsf{p}^{L-1}(i)=1$ for arbitrary leaf laminate $i$. Due to the absence of input rotation matrices for $(\tens{C}_1,\tens{C}_2)$, the material nodes $1\leq i\leq\nnodes$ that share the same leaf laminate also obtain the same effective rotation matrix. For instance, for the material nodes $3$ and $4$ contained in the leaf laminate $5$ in Fig. \ref{fig:dmn_laminate}, their effective rotation is
\begin{equation*}
    \tilde{\vec{R}}_5=\vec{R}(\vtheta_1)\vec{R}(\vtheta_2)\vec{R}(\vtheta_5).
\end{equation*}

Using these effective rotations on leaf laminates, similar to \eqref{eq:aij_3}, the DMN material frame orientation tensors can be computed for each phase $p$
\begin{equation} \label{eq:aij_DMN}
    \saijop^{(i)}_p(\vec{w},\vtheta)=\frac{\sum_{j\in\mathbb{I}_p}w_j\tilde{\vec{R}}_j^{(i)}\otimes\tilde{\vec{R}}_j^{(i)}}{\sum_{j\in\mathbb{I}_p} w_j},\quad p\in\{1,2\}.
\end{equation}
Compared to the DMN volume fraction \eqref{eq:dmn_vf}, the computation of DMN orientation tensors requires both DMN weights $\vec{w}$ and rotations $\vtheta$. Using the MI architecture \eqref{eq:physics_informed}, we propose the following orientation constraint through the definition of a loss function
\begin{equation} \label{eq:pinndmn_aij_loss}
    \mathcal{L}_{\saij}=\frac{1}{n_{\saij}}\sum_{p=1}^2\sum_{i=1}^{n_{\saij}}\norm{\saijop_p\bigl(\vec{w}(\vec{p}_i),\vtheta(\vec{p}_i)\bigr)-\saij_{\Omega,p}(\vec{p}_i)}^2,\quad \norm{\saij}^2=\sum_{i=1}^3\sum_{j=1}^3\sum_{k=1}^3 \left(a_{jk}^{(i)}\right)^2.
\end{equation}
In \eqref{eq:pinndmn_aij_loss}, $\vec{p}_i$ are the $n_{\saij}$ collocation points in the microstructural parametric space and $\saij_{\Omega,p}(\vec{p}_i)$ are the corresponding orientation tensors of the parameterized microstructure, for the phase $p$. The training strategy of \eqref{eq:pinndmn_aij_loss} will be described further in Sect. \ref{sec:training}.

\begin{figure}[htbp]
    \centering
    \includegraphics[width=0.8\textwidth]{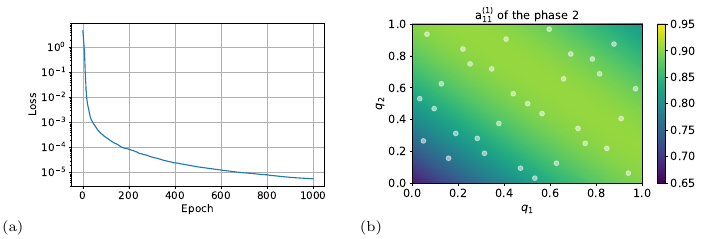}
    \caption{Orientation constraint prescribed on $\vec{w}(\vec{p})$ and $\vtheta(\vec{p})$: (a) minimization of the loss function \eqref{eq:pinndmn_aij_loss}; (b) material frame orientation tensor component $a^{(1)}_{11}$ for the 1st material axis $\vec{e}_1$ of the phase 2, at epoch 10 of training. The points in (b) represent the $\nnodes=2^5=32$ collocation points.} \label{fig:nn_rotation_training}
\end{figure}

With 5 DMN layers and the ReLU activation function, an example of the minimization of \eqref{eq:pinndmn_aij_loss} is shown in Fig. \ref{fig:nn_rotation_training}. The microstructure is parameterized by two geometrical parameters $\vec{p}=\vec{q}=(q_1,q_2)\in[0,1]^2$ that do not change $\vf$. Due to \eqref{eq:pwWpz}, the DMN weights do not vary with $\vec{p}$. We suppose that the material frame of both phases coincides with the global one, hence in \eqref{eq:pinndmn_aij_loss} the unidirectional orientation tensor \eqref{eq:saij_ud} is used as the target values. The decreasing loss history indicates that our MI architecture \eqref{eq:physics_informed} is capable of satisfying the orientation constraint.

At the early stage of the training, $a^{(1)}_{11}$ of the phase 2 still presents variations in the parametric space $(q_1,q_2)$. Its mean and standard deviation values at epoch $10^1$ and at epoch $10^3$ are indicated in Tab. \ref{tab:saij_1_11_mean_std}. After training, $a^{(1)}_{11}\approx 1$ becomes quasi-uniform as expected.
\begin{table}[htbp]
    \centering
    \begin{tabular}{lll} \toprule
        $a^{(1)}_{11}$  & Mean  & Standard deviation \\ \midrule
        At epoch $10^1$ & 0.866 & 0.0440             \\
        At epoch $10^3$ & 0.999 & 0.000686           \\ \bottomrule
    \end{tabular}
    \caption{Mean and standard deviation of $a^{(1)}_{11}$ of the phase 2 in the parametric space at epoch $10^1$ and at epoch $10^3$.}
    \label{tab:saij_1_11_mean_std}
\end{table}

\subsection{Offline training of MIpDMN} \label{sec:training}
As shown in Fig. \ref{fig:dmn_parametric}, the MIpDMN fitting parameters in \eqref{eq:physics_informed} are optimized \emph{jointly} using the linear elastic behavior data at different microstructural parameters. Each sample $s$ in the training data contains the homogenized linear elastic stiffness tensor $\CBar_s^\mathrm{FE}$ of a particular microstructure with parameters $\vec{p}_s$, given input linear elastic behaviors of both phases $(\tens{C}_1,\tens{C}_2)_s$, see Fig. \ref{fig:dmn_ferve}. Recall that $\CBar_s^\mathrm{FE}$ is in generally performed by computational homogenization like FE-RVE. The generation of such synthetic dataset requires sampling both in the input material space and in the parametric space of the microstructure
\begin{equation} \label{eq:sampling}
    (\tens{C}_1,\tens{C}_2)_s\in\mathbb{M},\quad \vec{p}_s=(\vf,\vec{q})_s\in\mathbb{P}\subset\mathbb{R}\times\mathbb{R}^q\quad\implies\CBar_s^\mathrm{FE}.
\end{equation}
In \eqref{eq:sampling}, $\mathbb{M}$ represents the discrete material sample set containing various input material properties of $(\tens{C}_1,\tens{C}_2)$, while $\mathbb{P}$ is the parametric sample set. In this work, these two samplings are performed \emph{independently} then combined using a \emph{Cartesian product}. For different $\vec{p}_s$, the \emph{same} $\mathbb{M}$ is used to evaluate the linear elastic homogenization behavior. With this Cartesian product approach, generalization error can be quantified unambiguously in the material sampling space and in the parametric space. To maintain the training cost manageable, in this work we limit the number of total samples (FE-RVE simulations) to be $\abs{\mathbb{P}}\times\abs{\mathbb{M}}=1500$. Hence, if more $\vec{p}_s$ are present, less $(\tens{C}_1,\tens{C}_2)_s$ will be sampled. For comparison, in the original DMN \cite{Liu:2019a} and the transfer-learning approach \cite{Liu:2019b}, they have fixed $\abs{\mathbb{M}}$ to be 500.

In \cite{Gajek:2021}, another strategy is proposed for short-fiber reinforced plastics with fixed volume fraction but arbitrary fiber orientations. The material samples set $\mathbb{M}$ is assigned to the parametric sample set $\mathbb{P}$ in a \emph{cyclic} fashion, resulting thus in a reduced number of $\abs{\mathbb{M}}$ samples. Since different material samples are used at different $\vec{p}_s$, it is important that no bias is introduced when distributing $\mathbb{M}$ into $\mathbb{P}$.

\paragraph{Material sampling} We follow the original material sampling method proposed by \cite{Liu:2019a}, assuming that $(\tens{C}_1,\tens{C}_2)$ are both orthotropic in their respective material frames. In total, $9+9=18$ material parameters are required to characterize their orthotropic elastic behaviors and $1$ additional scaling parameter is used to introduce contrasts in the elastic moduli between the two phases. In this work, Latin hypercube sampling \cite{Mckay:2000} is used to sample this 19-dimensional space.

\begin{figure}[htbp]
    \centering
    \includegraphics[width=0.8\textwidth]{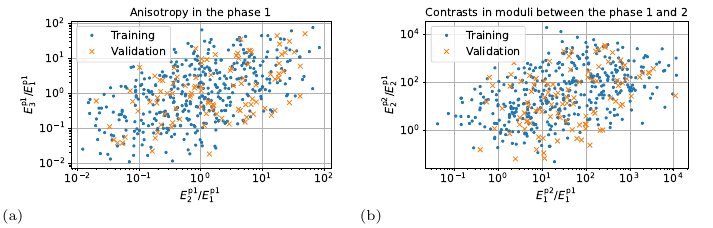}
    \caption{Material sampling for input orthotropic stiffness tensors $(\tens{C}_1,\tens{C}_2)$.} \label{fig:material_sampling}
\end{figure}

After sampling, the generated $(\tens{C}_1,\tens{C}_2)_i$ are randomly partitioned into a training set and a validation set. An example of material sampling is given in Fig. \ref{fig:material_sampling}, for the unidirectional fiber composite example in Sect. \ref{sec:ud}. Anisotropy in the phase 2 is similar to Fig. \ref{fig:material_sampling}(a) and is not shown. In this case, since the actual fibers are much stiffer than the matrix, the scaling parameter is adapted to generate appropriate contrasts in the synthetic elastic moduli between the two phases. From Fig. \ref{fig:material_sampling}(b), it can be seen that ratios between the Young's moduli range from $10^{-1}$ to $10^4$.

\paragraph{Sampling in the microstructural parametric space} Two different sampling procedures need to be performed in the microstructural parametric space. The first one defines the parameters values $\vec{p}_s\in\mathbb{P}$ in \eqref{eq:sampling} on which $\CBar_s^\mathrm{FE}$ is computed together with the previous material sampling $\mathbb{M}$. For different microstructures, different sampling strategy may be used to well cover the region of interest in the parametric space. For instance, in \cite{Gajek:2021}, the fiber orientation space is sampled by subdividing recursively the orientation triangle into sub-triangles. For microstructures depending solely on the volume fraction parameter \cite{Liu:2019b}, equidistant sampling can be used in the $\vf$-interval of interest. In general, all the microstructural parameters $\vec{p}=(\vf,\vec{q})$ are sampled jointly in order that less sampling points are needed to optimally fill the parametric space.

The other sampling concerns the physical constraints prescribed on \eqref{eq:physics_informed}. The volume fraction constraint \eqref{eq:pinndmn_vf_loss} and the orientation constraint \eqref{eq:pinndmn_aij_loss} are weakly enforced using collocation points in the parametric space. Since the costly computational homogenization is not carried out for these points, more collocation points can be used.

The volume fraction constraint \eqref{eq:pinndmn_vf_loss} requires a sampling of the $\vf$ parameter. A uniform sampling in $[0,1]$ is simply used to generate them: $\vf_i=i/(\nnodes-1)$, with $i=0,1,\ldots,\nnodes-1$. The number of collocation points is taken to be the number of DMN material nodes $\nnodes$. A numerical study not reported here shows that $\nnodes$ is sufficient to prescribe \eqref{eq:pinndmn_vf} with an acceptable accuracy, which is demonstrated by the example presented in Sect. \ref{sec:vf_constraint}.

The orientation constraint \eqref{eq:pinndmn_aij_loss} requires collocation points $\vec{p}_i$ in the parametric space, since it is prescribed both on DMN weights and rotations. Assume that each microstructural parameter is rescaled such that $\vec{p}\in[0,1]^{q+1}$. Sobol low-discrepancy sequence \cite{Sobol:1967} is used to generate these collocation points which presents good space filling properties. If internal bounds within the parameters are present, such as for the fiber orientation space \cite{Gajek:2021}, collocation points outside the feasible region are removed. The number of collocation points is also taken to be the number of DMN material nodes $\nnodes$. Sobol sequence is deterministic so that the same collocation points are used given the number of layers $L$. Numerical example shown in Sect. \ref{sec:training} shows that the orientation constraint can be well prescribed with this sampling strategy.

\paragraph{Loss function} Based on the previous material and parameter sampling, a total loss function $\mathcal{L}$ is defined as follows
\begin{equation} \label{eq:loss}
    \mathcal{L}=\frac{1}{\abs{\mathbb{P}}}\sum_{\mathbb{P}}\mathcal{L}_\vec{p}+\lambda_{\vf}\mathcal{L}_{\vf}+\lambda_{\saij}\mathcal{L}_{\saij},\qquad \mathcal{L}_\vec{p}=\frac{1}{\abs{\mathbb{M}}}\sum_{\mathbb{M}} e_i^2,\qquad e_i=\frac{\norm{\CBar_i^\mathrm{DMN}-\CBar_i^\mathrm{FE}}}{\norm{\CBar_i^\mathrm{FE}}}
\end{equation}
In \eqref{eq:loss}, $\mathcal{L}_{\vf}$ is the volume fraction constraint \eqref{eq:pinndmn_vf_loss} and $\mathcal{L}_{\saij}$ is the orientation constraint \eqref{eq:pinndmn_aij_loss}. Similar to \cite{Liu:2019a}, the data loss at a fixed microstructural parameter value $\mathcal{L}_\vec{p}$ is the mean squared error in the material sample set $\mathbb{M}$ comparing the DMN predictions and the computational homogenization ones, with the relative Frobenius norm on the stiffness tensors. In the total loss, $\mathcal{L}_\vec{p}$ is then averaged among different samples of the microstructural parameters.

The scaling factors $\lambda_{\vf}$ and $\lambda_{\saij}$ are \emph{symbolically} present in \eqref{eq:loss}. In the physics-informed machine learning community, they are introduced to balance the interplay between various data-based and PDE residual terms \cite{Henkes:2022}. In this work, we simply set $\lambda_{\vf}=\lambda_{\saij}=1$. Our numerical simulations indicate that both the data loss and the constraint terms are already of the same order $\mathcal{O}(1)$ initially. Furthermore, no improvement in terms of training, validation and test accuracy has been found by adjusting these scaling factors (for instance, by setting $\lambda_{\vf}=\lambda_{\saij}=10$). This may be related to the physics-based nature of DMN, the mathematically concise constraints \eqref{eq:pinndmn_vf_loss} and \eqref{eq:pinndmn_aij_loss} which are derivative-free, and the optimization algorithm described below.

In \cite{Liu:2019a,Gajek:2020}, a penalty term is also introduced to control the magnitude of $\vec{z}$ in \eqref{eq:w_from_z}. It is motivated by the fact that the ReLU activation function is unbounded and the DMN homogenization function \eqref{eq:DMN_recursive} is invariant with respect to the scaling $\vec{z}\mapsto k\vec{z}$ for $k>0$. In this work, we don't find it necessary to prescribe such constraints on our fitting parameters $(\vec{w}_0,\vec{w}_1)$.

\paragraph{Optimization algorithm} The total loss function \eqref{eq:loss} is minimized using gradient-based methods, similar to other DMN formulations \cite{Liu:2019,Gajek:2020,Wu:2021,Nguyen:2022}. Our MIpDMN architecture is implemented using PyTorch \cite{Paszke:2019}. The derivatives of the loss function \eqref{eq:loss} with respect to the fitting parameters \eqref{eq:physics_informed} can thus be easily computed with automatic differentiation. Single precision is used for offline training.

Our MIpDMN forward function $(\vec{p},\mathbb{C}_1,\mathbb{C}_2)\mapsto\CBar$ is vectorized over both the material sampling dimensions $(\mathbb{C}_1,\mathbb{C}_2)\in\mathbb{M}$ and the parameter sampling dimension $\vec{p}\in\mathbb{P}$, in order to achieve optimal parallel computational efficiency. In \cite{Liu:2019,Gajek:2021,Wu:2021,Nguyen:2022}, a mini-batch (or even only one sample) is randomly drawn from the whole dataset at each iteration of an epoch to introduce randomness in gradient descent and promote more frequent parameter updates. In the machine learning community, this is known to improve generalization capability \cite{Keskar:2017}. In this work however, a batch-learning approach is adopted, where the entire dataset is trained in a single batch at every epoch. Our simulation results indicate that the trained MIpDMN still generalizes well without significant overfitting, thanks to its physics-based nature. Furthermore, the training time can be reduced due to vectorization.

We have compared different first-order gradient-based optimizers provided in PyTorch, and have found that the resilient backpropagation (Rprop) algorithm \cite{Riedmiller:1993} works best in terms of convergence (loss values), stability (noise) and efficiency (time to train each epoch) in this batch-learning setting. Compared to other algorithms like Adam \cite{Kingma:2015} used in the DMN community \cite{Gajek:2021,Nguyen:2022a}, Rprop only considers the signs of gradients and each weight is updated independently using a dynamically adapted step size. For batch learning, Rprop with default hyperparameters outperforms Adam even if the hyperparameters of the latter are tuned \cite{Florescu:2018}. Hence, our MIpDMN is trained over 10000 epochs using Rprop with an initial learning rate (step-size) of $10^{-2}$.

\paragraph{Initialization} Depending on the activation function used in \eqref{eq:pwWpz}, different initialization can be used for $\vec{w}_0$. For ReLU, uniform distribution $\vec{w}_0\sim\mathcal{U}(0.2,0.8)$ as proposed by \cite{Liu:2019} is applied. For $\vec{w}_1$ which characterizes the dependence of the DMN weights on $\vf$, it is zero-initialized. Compared to a random initialization, numerical simulations demonstrate that it would result in better convergence of the loss function.

Regarding the DMN rotations in \eqref{eq:ptTpt}, similarly $\vec{\Theta}_1$ is zero-initialized. The dependence of $\vec{p}\mapsto\vtheta(\vec{p})$ is hence also learned from zero. For the constant part $\vtheta_0$, a random initialization on quaternions is used. Each component is initialized using the standard normal distribution, and then normalized to obtain a unit norm. We don't find it necessary to control the norm of each DMN rotation $\vtheta$ in \eqref{eq:ptTpt} through the use of an additional loss term in \eqref{eq:loss}.

Since initialization can have an impact on trained DMN parameters, training is in general repeated 20 times and the model with the least final loss value is chosen for further numerical investigations.

\section{Numerical examples} \label{sec:numerical}
In this section we will numerically evaluate our MIpDMN architecture on three parameterized microstructures. For the effective stiffness and thermal conductivity tensors, the FE-RVE simulation data are generated by imposing periodic boundary conditions \cite{Geers:2010} on the displacement or the temperature field. The effective CTE data are obtained by post-processing the thermal strains due to a uniform unit temperature increase, using \eqref{eq:thermoelastic_constitutive}. The finite element model is set up via the Abaqus FE-RVE plugin \cite{McLendon:2017}, then solved by Abaqus 2023 using 24 threads of Intel(R) Xeon(R) Gold 5220R CPU 2.20GHz. Training of MIpDMN is performed on an NVIDIA GeForce RTX 2080 Ti GPU card with single precision.

\subsection{Unidirectional fiber composite} \label{sec:ud}
\begin{figure}[htbp]
    \centering
    \includegraphics[width=0.95\textwidth]{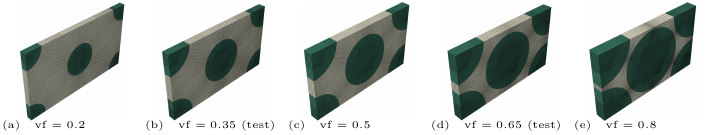}
    \caption{FE-RVE models for the unidirectional fiber composite with varying fiber volume fractions. The models with $\vf=0.2$, $\vf=0.5$ and $\vf=0.8$ are used to generate training dataset, while those with $\vf=0.35$ and $\vf=0.65$ are used to test interpolation accuracy.} \label{fig:hexa_meshes}
\end{figure}

We first consider a unidirectional fiber composite with varying fiber volume fractions. The 3-d finite element model is built with hexagonal fiber packing. In total, 5 FE-RVE models are constructed, see Fig. \ref{fig:hexa_meshes}. Three of them ($\vf=0.2$, $\vf=0.5$ and $\vf=0.8$) constitute the parametric sample set $\mathbb{P}$ and are used to generate training dataset, while the other two are used to test interpolation accuracy.

Material sampling $\mathbb{M}$ is performed using the method described in Sect. \ref{sec:training}. In total, the same 500 input orthotropic material properties are sampled for all the five microstructures, see Fig. \ref{fig:material_sampling}. Since the actual composite is made of a polypropylene matrix reinforced with glass fibers, the synthetic fiber stiffnesses $\mathbb{C}_2$ are also generated to be statistically higher than the matrix ones $\mathbb{C}_1$. The real linear elastic properties of the two isotropic phases are indicated in Tab. \ref{tab:hexa_real_properties}.
\begin{table}[htbp]
    \centering
    \begin{tabular}{lllll} \toprule
               & $E$ (MPa) & $\nu$ & $k$ (W/(m$\cdot$K)) & $\alpha$ ($\mathrm{K}^{-1}$) \\ \midrule
        Matrix & 3300      & 0.41  & 0.27                & $7\times 10^{-5}$            \\
        Fiber  & 72000     & 0.22  & 0.93                & $5\times 10^{-6}$            \\ \bottomrule
    \end{tabular}
    \caption{Real linear elastic properties, thermal conductivity and CTE of the two phases for the unidirectional fiber composite and the ellipsoidal inclusion composite.} \label{tab:hexa_real_properties}
\end{table}

For training microstructures $\vf=0.2$, $\vf=0.5$ and $\vf=0.8$, 400 of the 500 samples are used as training dataset, while the other 100 are reserved for validation. For the others $\vf=0.35$ and $\vf=0.65$, all the 500 samples are used for testing the interpolation accuracy. The FE-RVE models contain approximately $100\times 10^3$ to $200\times 10^3$ degrees of freedom and require up to 6 seconds for each run.

Since the volume fraction (of the fibers) is the only parameter of this microstructure, the MIpDMN architecture \eqref{eq:physics_informed} implies that the DMN weights $\vec{w}$ vary with $\vf$ according to \eqref{eq:pwWpz}, while the DMN rotations $\vtheta$ remain \emph{constant}. The fitting parameters are hence $(\vec{w}_0,\vec{w}_1,\vtheta_0)$. This is somehow similar to the ``naïve'' approach described in \cite{Liu:2019b}. Given a pre-trained base DMN with $\vec{w}^\mathrm{(b)}$ and corresponding volume fraction $\vf^\mathrm{(b)}$, the DMN weights at a new $\vf$ are scaled appropriately based on the volume fraction
\begin{equation} \label{eq:naive}
    w_i=\begin{cases}
        \dfrac{1-\vf}{1-\vf^\mathrm{(b)}}w^\mathrm{(b)}_i & i\in\mathbb{I}_1, \\
        \dfrac{\vf}{\vf^\mathrm{(b)}}w^\mathrm{(b)}_i     & i\in\mathbb{I}_2. \\
    \end{cases}
\end{equation}
The DMN rotations remain unchanged. Compared to \eqref{eq:naive}, the $\vf$-dependence of $\vec{w}$ in our MIpDMN is not directly prescribed but learned from data.

\begin{figure}[htbp]
    \centering
    \includegraphics[width=0.4\textwidth]{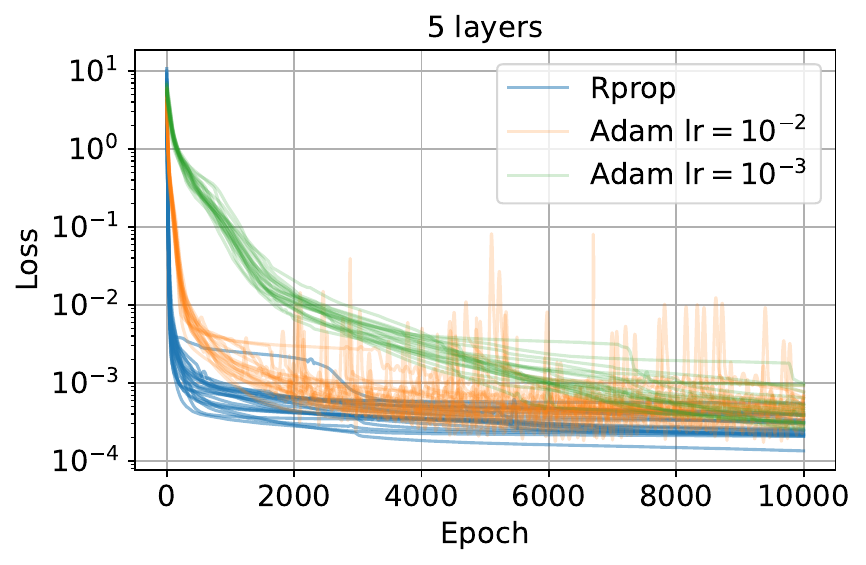}
    \caption{Loss history with 5 DMN layers with the Rprop and the Adam optimizers using two learning rates (lr).} \label{fig:hexa_loss_adam_rprop}
\end{figure}

Training of our MIpDMN is conducted following Sect. \ref{sec:training}, with the two physical constraints \eqref{eq:pinndmn_vf_loss} and \eqref{eq:pinndmn_aij_loss}. The ReLU activation function is used. Since the material frames of both phases coincide with the global one, the unidirectional material frame orientation tensor \eqref{eq:saij_ud} is used as the target value for the orientation constraint. With 5 DMN layers, we first compare Rprop with the frequently used Adam optimizer in terms of loss history, in Fig. \ref{fig:hexa_loss_adam_rprop}. A total of 20 trainings are realized using random initialization for each optimizer. For Adam, two (initial) learning rates are used, which are common values for DMN training \cite{Gajek:2021,Nguyen:2022a}. With a larger $\mathrm{lr}=10^{-2}$, Adam results in a much more noisy loss history which could be solved by learning rate decay \cite{Gajek:2021,Nguyen:2022a}, at the cost of introducing more hyperparameters. For both learning rates, Adam converges slower than Rprop with no improvement in the training accuracy. Similar to the finding of \cite{Florescu:2018}, our numerical simulations indicate that Rprop with no hyperparameter tuning outperforms Adam for the batch-training of MIpDMN.

\begin{figure}[htbp]
    \centering
    \includegraphics[width=0.8\textwidth]{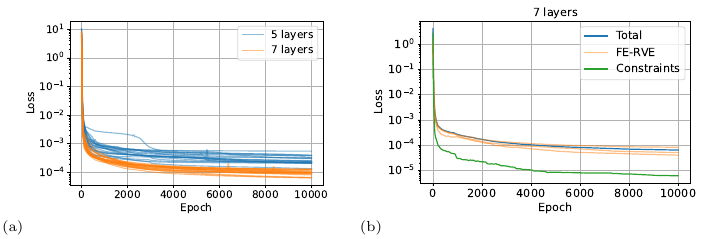}
    \caption{MIpDMN training for the unidirectional fiber composite: (a) loss histories with 5 and 7 DMN layers; (b) partition of the total loss into the FE-RVE data part and the physical constraints part.} \label{fig:hexa_loss}
\end{figure}

Similar to the original DMN formulations \cite{Liu:2019}, adding more layers also increases the expressive power of MIpDMN. From Fig. \ref{fig:hexa_loss}(a), the final loss values with 7 layers are lower than those obtained with 5 layers. The total loss can be partitioned into two parts according to \eqref{eq:loss}: the FE-RVE data part and the physical constraints part. Their respective history is shown in Fig. \ref{fig:hexa_loss}(b) for the training realization with the least final loss value, for 7 layers. Each of the three ``FE-RVE'' curves represents one particular training microstructure $\vf=0.2$, $\vf=0.5$ or $\vf=0.8$, and is monotonically decreasing simultaneously with similar values. The physical constraints part also converges well and is lower than the FE-RVE part by an order of magnitude. Among the 20 training realizations, the median training time for 5 or 7 layers is only respectively 570 s and 786 s. Compared to the training times (several hours) reported in \cite{Liu:2019a,Gajek:2021}, we believe that our batch-learning approach along with the vectorization of the forward function help to speed up training significantly. In the sequel, we will report the results obtained with 5 layers which provides satisfactory accuracy.

\begin{figure}[htbp]
    \centering
    \includegraphics[width=0.4\textwidth]{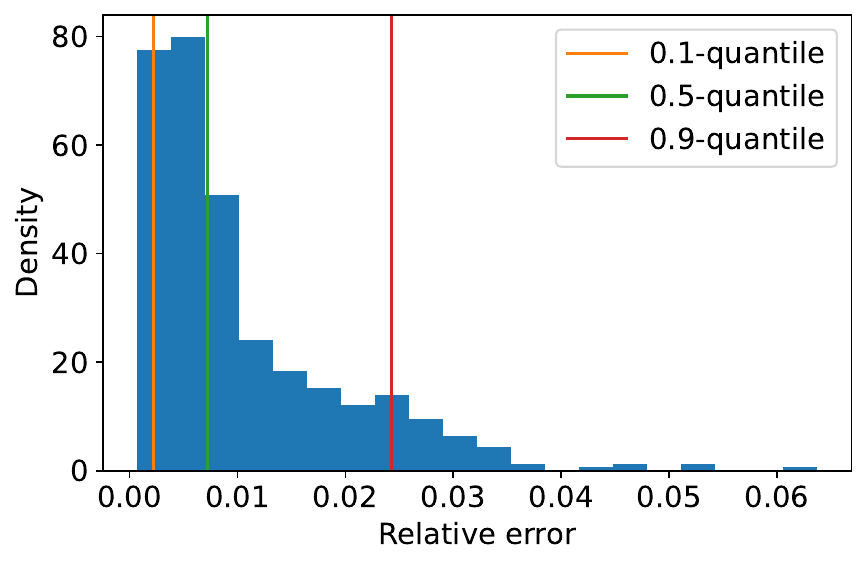}
    \caption{Statistical distribution of the relative error $e_i$ for the microstructure with $\vf=0.5$.} \label{fig:hexa_quantiles}
\end{figure}

In Fig. \ref{fig:hexa_quantiles}, the statistical distribution of the relative Frobenius-norm error $e_i$ in \eqref{eq:loss} between DMN predictions and FE-RVE ones on each material sample is shown for the microstructure with $\vf=0.5$. The 0.1, 0.5 (median) and 0.9-quantiles are also indicated. Even though the maximum relative error can reach 6\% for some localized samples, the error is less than 2.4\% for 90\% of them. We believe that these quantiles are less sensitive to outliers and hence more appropriate for comparing different models.

\begin{figure}[htbp]
    \centering
    \includegraphics[width=0.8\textwidth]{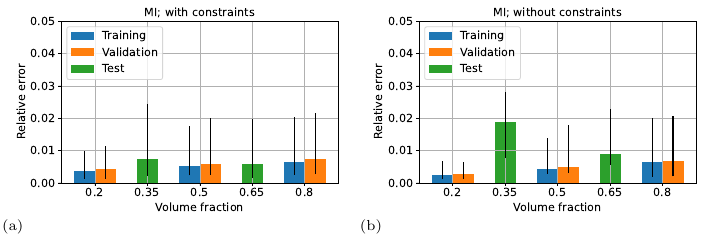}
    \caption{Training, validation and test errors of MIpDMN at different volume fractions: (a) with physical constraints; (b) without physical constraints.} \label{fig:hexa_training_error_pinndmn}
\end{figure}

We first analyze the influence of physical constraints on the MI architecture \eqref{eq:physics_informed}. The training, validation and test errors are compared in Fig. \ref{fig:hexa_training_error_pinndmn} using their respective 0.1, 0.5 (median) and 0.9-quantiles at different volume fractions. The MIpDMN model without physical constraints is trained similarly with 20 realizations and the one with the least final loss value is chosen. Validation error refers to the relative error computed on the validation dataset composed of 100 material samples. Overfitting is not observed for both cases, since the validation errors are comparable to the training errors. When the physical constraints are not included, the training errors at $\vf=0.5$ and $\vf=0.8$ are similar or only slightly lower at $\vf=0.2$ than the case when they are considered. However, the inclusion of the physical constraints reduces the interpolative test errors at $\vf=0.35$ and $\vf=0.65$ for previously unseen microstructures. The physical constraints may hence improve generalization ability of MIpDMN.

\begin{figure}[htbp]
    \centering
    \includegraphics[width=0.8\textwidth]{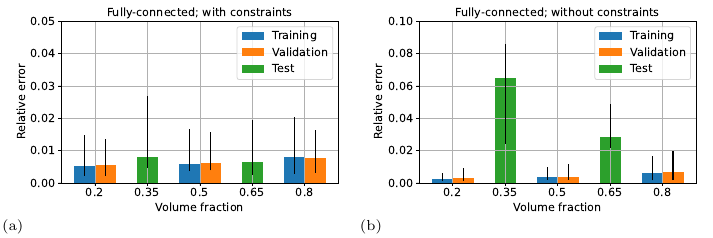}
    \caption{Training, validation and test errors of a fully-connected parametric DMN at different volume fractions: (a) with physical constraints; (b) without physical constraints.} \label{fig:hexa_training_error_fully}
\end{figure}

In Fig. \ref{fig:hexa_training_error_fully}, we analyze the influence of the physical constraints on the fully-connected architecture \eqref{eq:fully_connected}. Now, the DMN rotations also become a function of the volume fraction, leading to an additional fitting parameter $\vec{\Theta}_1$ during training. The physical constraints reduce overfitting at the data points. By comparing Fig. \ref{fig:hexa_training_error_pinndmn}(a) and Fig. \ref{fig:hexa_training_error_fully}(a), we observe similar training, validation and test errors between the MI and the fully-connected architectures, when the physical constraints are included. The MI architecture \eqref{eq:physics_informed} hence demonstrates similar expressive power compared to the fully-connected one, even with a constant DMN rotations vector $\vtheta_0$ for different volume fractions. When the physical constraints are not considered, the interpolative test errors become significantly larger (even the scale of the $y$-axis needs to be changed) in Fig. \ref{fig:hexa_training_error_fully}(b) for the fully-connected architecture.  Comparing Fig. \ref{fig:hexa_training_error_pinndmn}(b) and Fig. \ref{fig:hexa_training_error_fully}(b), we conclude that the MI architecture itself may help to improve generalization capability.

\begin{figure}[htbp]
    \centering
    \includegraphics[width=0.8\textwidth]{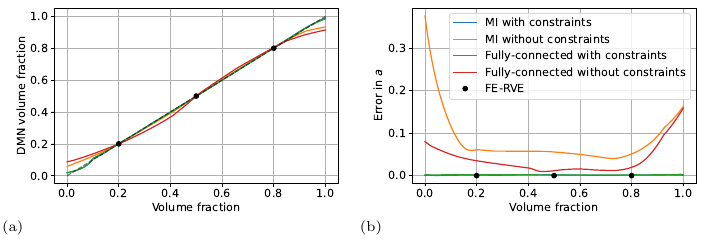}
    \caption{Physical constraints verification for the micromechanics-informed (MI) and the fully-connected architectures: (a) DMN volume fraction prediction; (b) error on the material frame orientation tensors.} \label{fig:hexa_constraint_with_without}
\end{figure}

In Fig. \ref{fig:hexa_constraint_with_without}(a), the DMN volume fraction prediction \eqref{eq:dmn_vf} is presented as a function of the microstructure volume fraction. For all cases, DMN recovers well the volume fractions $0.2$, $0.5$ and $0.8$ at the FE-RVE data points. The microstructure morphologies are hence well learned by DMN using linear elastic data \cite{Liu:2019,Gajek:2020}. However, away from these three data points, the DMN volume fraction prediction may differ from the actual microstructure one. At $\vf=0.35$, when \eqref{eq:pinndmn_vf_loss} is explicitly included during training, the relative errors $\abs{\vf_{\DMN}-\vf_\Omega}/\vf_\Omega$ are smaller than 0.4\% for the two parametric DMN architectures. However, when it is not the case, we obtain 2\% with the MI architecture and even 9\% with the fully-connected one. This may explain the higher interpolative testing errors at $\vf=0.35$ for these two models in Fig. \ref{fig:hexa_training_error_pinndmn}(b) and Fig. \ref{fig:hexa_training_error_fully}(b). The straight line \eqref{eq:pinndmn_vf} can only be well recovered when the volume fraction constraint \eqref{eq:pinndmn_vf_loss} is explicitly included.

Similarly, the error on the material frame orientation tensors $\norm{\saijop_p(\vf_i)-\saij_{\Omega,p}(\vf_i)}$ using the norm defined in \eqref{eq:pinndmn_aij_loss} is presented in Fig. \ref{fig:hexa_constraint_with_without}(b). The errors on the fibers and on the matrix are summed together. When \eqref{eq:pinndmn_aij_loss} is not included during training, deviation from the theoretical unidirectional material frame orientation tensor \eqref{eq:saij_ud} can be observed, even though the error is well bounded between $[0.2, 0.8]$. Similar to Fig. \ref{fig:hexa_constraint_with_without}(a), physical constraint errors increase significantly outside the training domain. The inclusion of \eqref{eq:pinndmn_aij_loss} may hence improve the generalization capability in the whole parametric space.

\begin{figure}[htbp]
    \centering
    \includegraphics[width=0.4\textwidth]{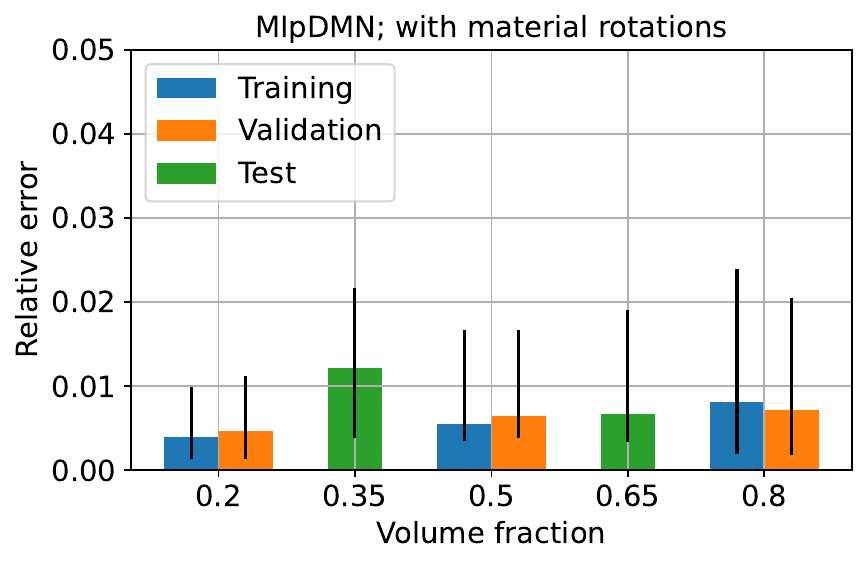}
    \caption{Training, validation and test errors of MIpDMN with additional $\nnodes=2^L=32$ material rotations at different volume fractions.} \label{fig:hexa_training_error_pinndmn_local_rotations}
\end{figure}

In this work, the rotation matrix for $(\tens{C}_1,\tens{C}_2)$ of each of the $\nnodes=2^L$ material nodes on the input layer is omitted compared to the original formulation \cite{Liu:2019,Liu:2019a}. In Fig. \ref{fig:hexa_training_error_pinndmn_local_rotations}, the training, validation and test errors are shown when they are instead included. Similarly, the training process is realized 20 times with random initialization and the model with the least final loss value is chosen. Compared to Fig. \ref{fig:hexa_training_error_pinndmn}(a), the inclusion of such material rotations does not increase the expressive power of MIpDMN. This could be partially due to the fact that the microstructure being considered does not contain local material orientation.

\begin{figure}[htbp]
    \centering
    \includegraphics[width=0.4\textwidth]{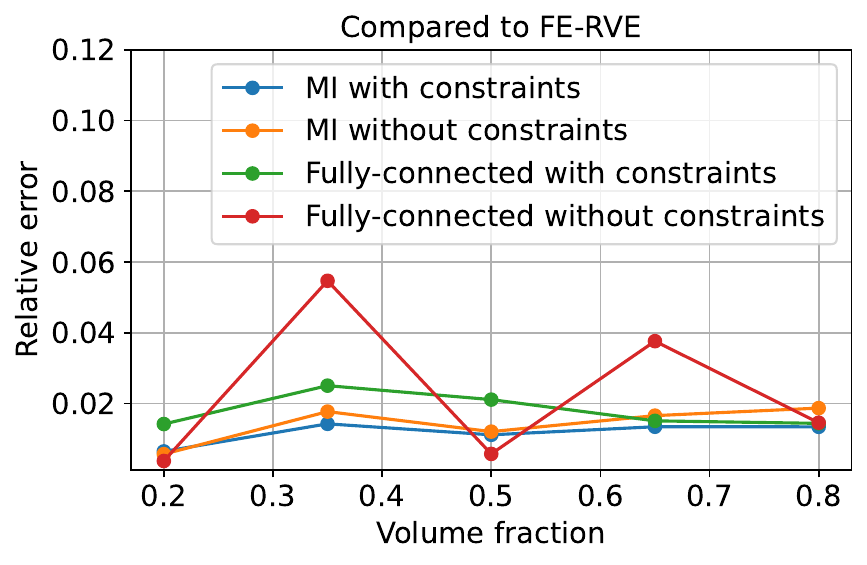}
    \caption{Relative error using the real properties between the DMN predictions and the FE-RVE results for two parametric DMN architectures with or without the physical constraints.} \label{fig:hexa_real_error_pinndmn_fully}
\end{figure}

Using the real linear elastic properties of the two phases in Tab. \ref{tab:hexa_real_properties}, the relative errors between the DMN predictions and the FE-RVE results for two parametric DMN architectures and with or without the physical constraints are shown in Fig. \ref{fig:hexa_real_error_pinndmn_fully}. The conclusions drawn from Fig. \ref{fig:hexa_training_error_pinndmn} and Fig. \ref{fig:hexa_training_error_fully} are recovered. The MI architecture and the physical constraints both help to improve generalization ability of the parametric DMN model. Using MIpDMN with the physical constraints, the relative errors are less than 1.5\% among the 5 microstructures being considered.

\begin{figure}[htbp]
    \centering
    \includegraphics[width=0.95\textwidth]{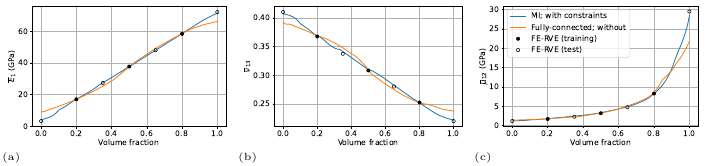}
    \caption{Homogenized elastic moduli prediction as a function of the fiber volume fraction obtained by parametric DMN models: (a) longitudinal Young's modulus $\overline{E}_1$; (b) transverse Poisson ratio $\overline{\nu}_{13}$; (c) transverse shear modulus $\overline{\mu}_{12}$. The theoretical values at $\vf=0$ and $\vf=1$ are also indicated as test data.} \label{fig:hexa_moduli_pinndmn_fully}
\end{figure}

The homogenized elastic property prediction is presented in Fig. \ref{fig:hexa_moduli_pinndmn_fully} in the \emph{whole} parametric space. The general nonlinear influence of the volume fraction is well captured by these two parametric DMN models. The use of the MI architecture as well as the physical constraints helps to improve generalization capabilities.

\begin{figure}[htbp]
    \centering
    \includegraphics[width=0.8\textwidth]{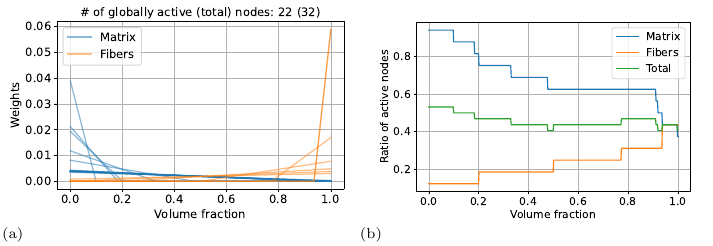}
    \caption{MIpDMN: (a) DMN weights evolution with the volume fraction; (b) variation of the ratio of active DMN nodes} \label{fig:hexa_n_dof_active_pinndmn}
\end{figure}

In Fig. \ref{fig:hexa_n_dof_active_pinndmn}(a), the MI functional dependence of the DMN weights \eqref{eq:pwWpz} is illustrated. The MIpDMN architecture with the physical constraints is used. Globally, 22 of the 32 material nodes are active. With the increasing fiber volume fraction, DMN weights of the matrix decrease \emph{gradually} and \emph{individually}, while those of the fiber increase at the same time. This is also reflected in Fig. \ref{fig:hexa_n_dof_active_pinndmn}(b), where the ratio of the active DMN nodes is presented for both phases. Even though the ratios for both phases gradually increase or decrease, the total number of active nodes does not vary much. This ensures the expressive power of MIpDMN in the whole parametric space.

\begin{figure}[htbp]
    \centering
    \includegraphics[width=0.8\textwidth]{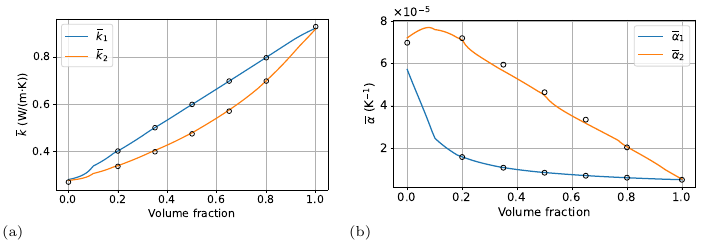}
    \caption{(a) Effective thermal conductivity prediction for the longitudinal $\overline{k}_1$ and transverse $\overline{k}_2$ components; (b) effective CTE prediction. The FE-RVE results and the theoretical values for $\vf=0$ and $\vf=1$ are indicated by circles.} \label{fig:hexa_vf_rev_k_alphaBar}
\end{figure}

Using the proposed MIpDMN architecture and the fitted parameters trained previously on isothermal linear elastic data, we will now predict the effective thermal conductivity $\kBar$ and the effective CTE $\alphaBar$ with the redefinitions of the laminate homogenization function $\Lam$ in Sect. \ref{sec:multiphi}. The constituent properties used in this online prediction test are provided in Tab. \ref{tab:hexa_real_properties}. In Fig. \ref{fig:hexa_vf_rev_k_alphaBar}, the MIpDMN predictions are compared with the FE-RVE results as a function of the volume fraction. The nonlinear $\vf$-dependences of $\kBar$ and $\alphaBar$ are well captured by MIpDMN. An excellent agreement is obtained in the whole interval, even though the longitudinal $\overline{\alpha}_1$ is slightly underestimated near $\vf\approx 0$. Using a similar relative Frobenius-norm error $e_i$ in \eqref{eq:loss} for second-order tensors, the maximum error among these 5 FE-RVE data points is 0.85\% for $\kBar$ at $\vf=0.35$, and 4.5\% for $\alphaBar$ also at $\vf=0.35$. This is a remarkable result knowing that our MIpDMN is only trained using isothermal linear elastic data. This demonstrates that DMN learns the microstructure \emph{per se}, and not a particular physics property in particular.

\subsubsection*{Comparison with transfer-learning based interpolative DMN}
We follow the procedure described in \cite{Liu:2019b} and the microstructure with the least fiber volume fraction is used as the starting point in the training sequence $\vf_0=0.2\to\vf_1=0.5\to\vf_2=0.8$. Subsequent training is initialized using \eqref{eq:naive} based on the previously trained DMN. As for our parametric DMN models, the transfer learning process is repeated 20 times with random initialization when training the base DMN model. The realization with the least final total loss value $\mathcal{L}_\mathrm{tot}=\mathcal{L}_0+\mathcal{L}_1+\mathcal{L}_2$ is then chosen for further analysis. For comparison with previous MIpDMN results, 5 DMN layers are used.

\begin{figure}[htbp]
    \centering
    \includegraphics[width=0.8\textwidth]{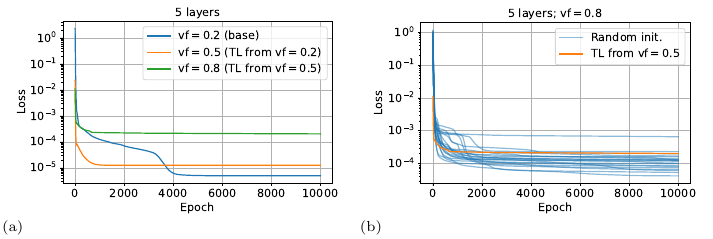}
    \caption{(a) Loss histories for different volume fractions using transfer learning; (b) loss history at $\vf_2=0.8$ obtained using random initialization and from transfer learning.} \label{fig:hexa_loss_transfer}
\end{figure}

In Fig. \ref{fig:hexa_loss_transfer}(a), loss histories obtained with transfer learning are presented. Transfer learning indeed accelerates convergence. For $\vf=0.5$ and $\vf=0.8$, convergence is achieved within the first 1000 epochs. However, the final loss value becomes an order of magnitude larger at $\vf=0.8$ compared to the base DMN. In \cite{Liu:2019b}, there is also a slight increase in the final loss values at the end of the transfer learning process. For $\vf=0.8$, DMN is also trained using random initialization. The loss histories are compared with that obtained from transfer learning. A reduction of the expressive power of DMN is hence observed with transfer learning.

\begin{figure}[htbp]
    \centering
    \includegraphics[width=0.8\textwidth]{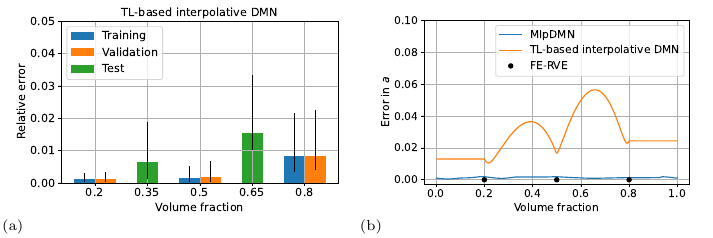}
    \caption{(a) Transfer-learning based interpolative DMN: training, validation and test errors at different volume fractions; (b) error on the material frame orientation tensors obtained by MIpDMN with the physical constraints and by the transfer-learning approach.} \label{fig:hexa_training_error_interpolative_aij_constraint_interpolative_revised}
\end{figure}

The training, validation and test errors obtained with this transfer-learning based interpolative DMN model are presented in Fig. \ref{fig:hexa_training_error_interpolative_aij_constraint_interpolative_revised}(a). Compared to the results obtained with MIpDMN (with the physical constraints) in Fig. \ref{fig:hexa_training_error_pinndmn}(a), the errors at the training points $\vf=0.2$ and $\vf=0.5$ are lower with this transfer learning approach. However, due to the reduction of the expressive power, the errors at $\vf=0.8$ become comparable to those with MIpDMN. The test errors are significantly larger than the training and the validation ones, especially at $\vf=0.65$. The transfer-learning may hence tend to overfit at the data points.

Even though it is not proposed in \cite{Liu:2019b}, we can also use \eqref{eq:naive} to extrapolate DMN weights outside the training domain. In this case, the DMN weights at $\vf=0$ are hence extrapolated from the base DMN at $\vf=0.2$, while the DMN trained by transfer-learning at $\vf=0.8$ can be used to obtain the DMN weights at $\vf=1$. Since \eqref{eq:naive} respects the actual volume fraction \eqref{eq:dmn_vf}, the volume fraction constraint $\vf_{\DMN}=\vf_{\Omega}$ is satisfied at all $\vf$ when combined with piecewise linear interpolation. However, it is not the case for the orientation constraint. In Fig. \ref{fig:hexa_training_error_interpolative_aij_constraint_interpolative_revised}(b), the error on the material frame orientation tensors is computed in the whole parametric space. Compared to MIpDMN, the error is much larger with the transfer-learning approach and also increases with interpolation. This may explain the larger test errors in the transfer-learning model in Fig. \ref{fig:hexa_training_error_interpolative_aij_constraint_interpolative_revised}(a).

\begin{figure}[htbp]
    \centering
    \includegraphics[width=0.4\textwidth]{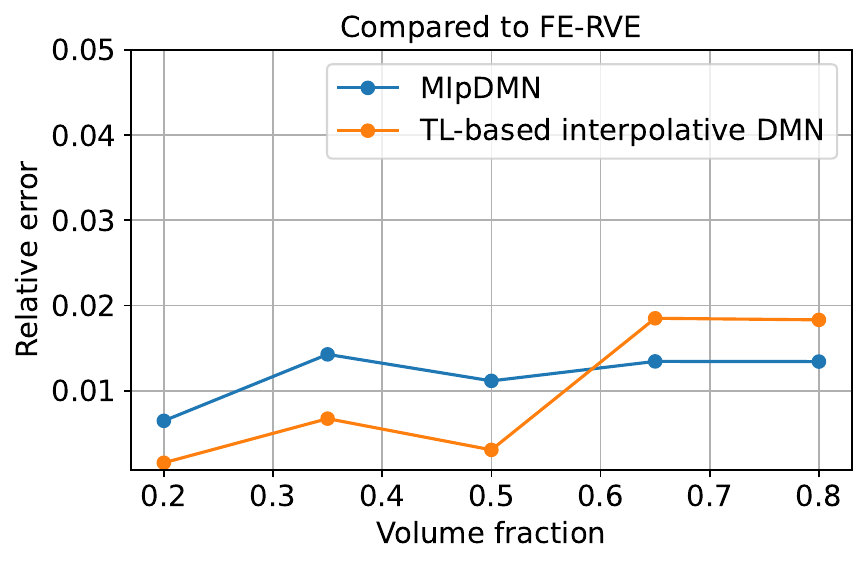}
    \caption{Relative error using real properties between the DMN predictions and the FE-RVE results obtained by MIpDMN with the physical constraints and by the transfer-learning (TL) approach.} \label{fig:hexa_real_error_pinndmn_interpolative}
\end{figure}

The relative errors using the real linear elastic properties of the two phases in Tab. \ref{tab:hexa_real_properties} are shown in Fig. \ref{fig:hexa_real_error_pinndmn_interpolative} for the MIpDMN model and the transfer-learning approach. Even though the latter is more accurate for the first 3 volume fractions, it becomes less accurate for the last 2 volume fractions. The relative errors obtained with MIpDMN, on the contrary, are more uniform in the parametric space.

\begin{figure}[htbp]
    \centering
    \includegraphics[width=0.8\textwidth]{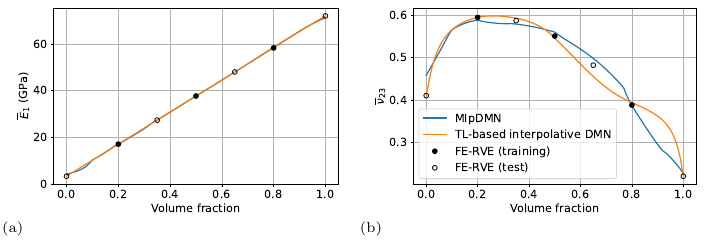}
    \caption{Homogenized elastic moduli prediction as a function of the fiber volume fraction obtained by MIpDMN and the transfer-learning (TL) approach: (a) longitudinal Young's modulus $\overline{E}_1$; (b) in-plane Poisson ratio $\overline{\nu}_{23}$.} \label{fig:hexa_moduli_pinndmn_interpolative}
\end{figure}

The elastic moduli prediction is also compared in Fig. \ref{fig:hexa_moduli_pinndmn_interpolative} between these two approaches. Both models are able to capture the nonlinear influence of the volume fraction, especially on the in-plane Poisson ratio $\nu_{23}$. While a perfect agreement is found for $\overline{E}_1$, the transfer-learning approach seems to slightly overfit the $\nu_{23}$ prediction between $0.5$ and $1.0$.

\begin{figure}[htbp]
    \centering
    \includegraphics[width=0.8\textwidth]{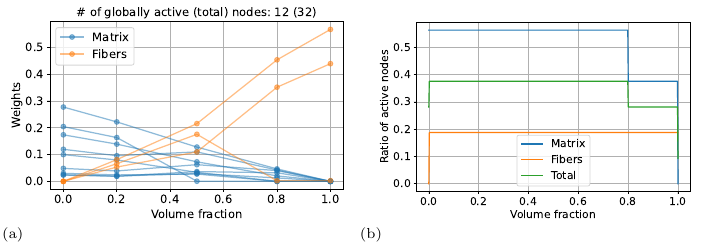}
    \caption{Transfer-learning based interpolative DMN: (a) DMN weights evolution with the volume fraction; (b) variation of the ratio of active DMN nodes.} \label{fig:hexa_n_dof_active_interpolative}
\end{figure}

Similar to Fig. \ref{fig:hexa_n_dof_active_pinndmn}, the DMN weights evolution in the parametric space is also shown in Fig. \ref{fig:hexa_n_dof_active_interpolative} for the transfer-learning approach. Globally, 12 of the 32 material nodes are active. Due to transfer-learning and the ReLU activation function, the number of active nodes can only decrease \cite{Liu:2019b}. While 12 material nodes are active for $\vf=0.2$ and $\vf=0.5$, only 9 active nodes are present for $\vf=0.8$. Compared to MIpDMN, this gradual decrease of the active DMN material nodes may lead to a reduced expressive power. The functional dependence on $\vf$ is mainly realized by adapting weights values at each transfer-learning interpolation point. There is no gradual change in the ratios of active nodes for intermediate volume fractions.

\subsection{Woven composite} \label{sec:woven}
\begin{figure}[htbp]
    \centering
    \includegraphics[width=0.95\textwidth]{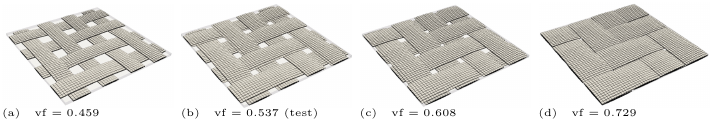}
    \caption{FE-RVE models for the woven composite with different tow volume fractions. The models with $\vf=0.459$, $\vf=0.608$ and $\vf=0.729$ are used to generate the training dataset, while $\vf=0.537$ is used to test interpolation accuracy.} \label{fig:woven_meshes}
\end{figure}

We now consider the $2\times 2$ twill woven composites with varying tow volume fractions. Compared to the unidirectional fiber composite, now local material orientation is present for the yarns, illustrated in Fig. \ref{fig:woven_e1e2e3}. The 3-d finite element model is constructed using TexGen \cite{Brown:2021} with 4 volume fraction values for the tows, see Fig. \ref{fig:woven_meshes}. Three of them ($\vf=0.459$, $\vf=0.608$ and $\vf=0.729$) constitute the parametric sample set $\mathbb{P}$ and are used to generate the training dataset, while $\vf=0.537$ is used to test interpolation accuracy. The FE-RVE model contains 75000 voxel elements and requires approximately 11 seconds for each run.

The real linear elastic properties of the two phases are adapted from \cite{Liu:2019a} and can be found in Tab. \ref{tab:woven_real_properties}. The matrix is isotropic while the carbon fiber tows are assumed to be transversely isotropic in the local material frames. Compared to \cite{Wu:2021}, the transverse Young's modulus $E_2$ of the tows is modified to satisfy transverse isotropy. A similar material sampling $\mathbb{M}$ compared to Fig. \ref{fig:material_sampling} is performed to generate 500 input orthotropic material properties. Given the high moduli contrasts in the real properties, the synthetic yarn stiffnesses $\mathbb{C}_2$ are also generated to be statistically higher than the matrix ones $\mathbb{C}_1$. For the training microstructures $\vf=0.459$, $\vf=0.608$ and $\vf=0.729$, 400 of the 500 samples are used as the training dataset, while the other 100 are reserved for validation. For $\vf=0.537$, all the 500 samples are used for testing the interpolation accuracy.
\begin{table}[htbp]
    \centering
    \begin{tabular}{llllll} \toprule
                           & $E$ (GPa) & $\nu$ & $\mu$ (GPa) & $k$ (W/(m$\cdot$K)) & $\alpha$ ($\mathrm{K}^{-1}$) \\ \midrule
        Matrix             & 3.8       & 0.387 & 1.37        & 0.3                 & $7\times 10^{-5}$            \\
        Tow (longitudinal) & 78.8      & 0.35  & 2.39        & 1                   & $4\times 10^{-6}$            \\
        Tow (transverse)   & 6.24      & 0.6   & 1.95        & 0.4                 & $4\times 10^{-5}$            \\
        \bottomrule
    \end{tabular}
    \caption{Real linear elastic properties, thermal conductivity and CTE of the two phases for the woven composite.} \label{tab:woven_real_properties}
\end{table}

MIpDMN is trained following Sect. \ref{sec:training} with the two physical constraints \eqref{eq:pinndmn_vf_loss} and \eqref{eq:pinndmn_aij_loss} and the ReLU activation function. In \eqref{eq:pinndmn_aij_loss}, the unidirectional material frame orientation tensor \eqref{eq:saij_ud} is defined as the target for the matrix, since its material frame coincides with the global one. For the tows, we use the ``planar isotropic'' tensor \eqref{eq:saij_tow} which characterizes the statistical spatial orientation of the yarns in the FE-RVE model. As before, the training process is realized 20 times with random initialization and the one with the least final loss value is chosen for further investigations. The median training time is only 800 s for 7 layers and 1176 s for 9 layers.

\begin{figure}[htbp]
    \centering
    \includegraphics[width=0.8\textwidth]{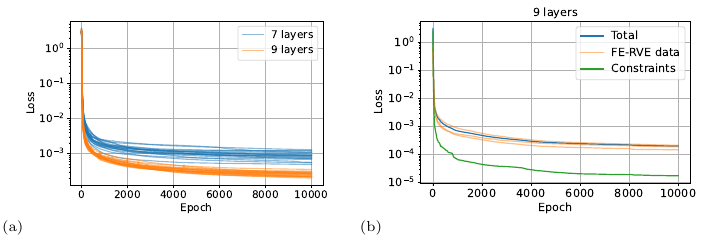}
    \caption{MIpDMN training for the woven composite: (a) loss histories with 7 and 9 DMN layers; (b) partition of the total loss into the FE-RVE data part and the physical constraints part.} \label{fig:woven_loss}
\end{figure}

In Fig. \ref{fig:woven_loss}, the loss histories are compared for 7 and 9 DMN layers. As for the unidirectional fiber composite, an increased expressive power is observed with more layers, leading to lower (final) loss values. The FE-RVE data and the physical constraints parts are monotonically decreasing. The physical constraints \eqref{eq:pinndmn_vf_loss} and \eqref{eq:pinndmn_aij_loss} are well satisfied since the corresponding loss value is approaching $10^{-5}$ with 9 layers. In the sequel, we will report the results using 7 layers which provides satisfactory accuracy.

\begin{figure}[htbp]
    \centering
    \includegraphics[width=0.8\textwidth]{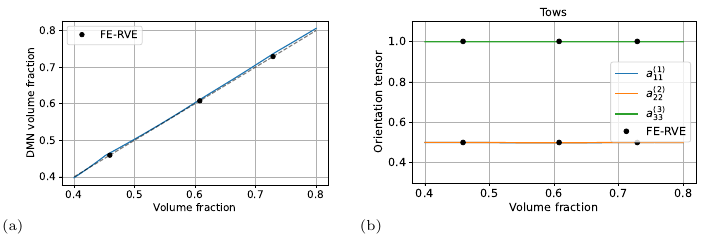}
    \caption{Verification of the physical constraints for the woven composite: (a) volume fraction; (b) material frame orientation tensors for the tows.} \label{fig:woven_constraints}
\end{figure}

The volume fraction and the material frame orientation tensors for the tows are compared with their prescribed values in Fig. \ref{fig:woven_constraints}. The DMN volume fraction matches the FE-RVE data points and agrees well with the theoretical straight line even when evaluated outside the training region. In Fig. \ref{fig:woven_constraints}(b), the components $a_{ii}^{(i)}$ of the DMN material orientation tensors are shown. An excellent agreement is also obtained. Even without the additional rotations on the input layer, our DMN architecture is capable of learning local material orientation present in the microstructure.

\begin{figure}[htbp]
    \centering
    \includegraphics[width=0.4\textwidth]{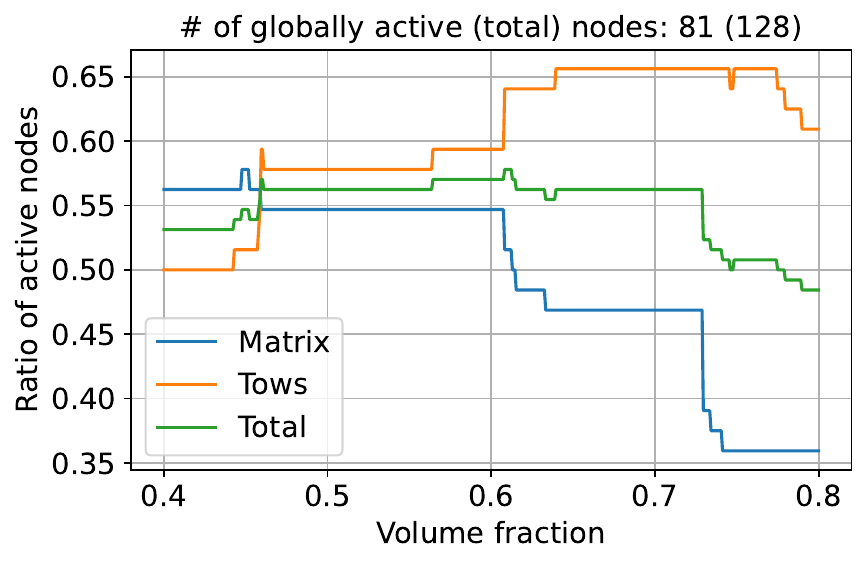}
    \caption{Variation of the ratio of active DMN nodes with varying volume fraction.} \label{fig:woven_n_dof_active}
\end{figure}

In total, 81 DMN material nodes are globally active in the parametric interval $[0.4, 0.8]$, as shown in Fig. \ref{fig:woven_n_dof_active}. As in Fig. \ref{fig:hexa_n_dof_active_pinndmn}, the number of active nodes gradually decreases from 55\% to 35\% for the matrix while that of the tows increases from 50\% to 65\%. The overall ratio of the active nodes remains approximately 55\% within the training region $[0.459, 0.729]$. This is believed to ensure sufficient expressive power for MIpDMN.

\begin{figure}[htbp]
    \centering
    \includegraphics[width=0.8\textwidth]{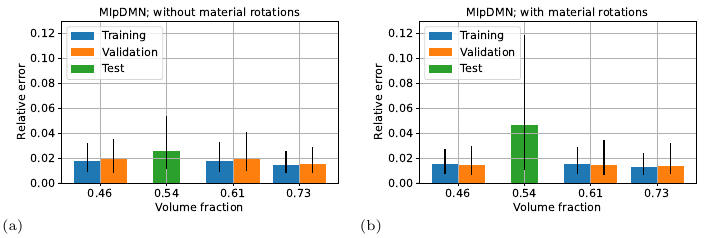}
    \caption{Training, validation and test errors of MIpDMN at different volume fractions: (a) without material rotations; (b) with additional $2^7=128$ material rotations.} \label{fig:woven_training_error}
\end{figure}

In Fig. \ref{fig:woven_training_error}(a), the training, validation and test errors are computed at different volume fractions for MIpDMN without material rotations. The median errors are approximately 2\% for all the four microstructures. As in Fig. \ref{fig:hexa_training_error_pinndmn}(a), the test error is only slightly larger compared to the training ones. In Fig. \ref{fig:woven_training_error}(b), the results are reported for MIpDMN with the additional $2^7=128$ material rotations. The training is also realized 20 times and the model with the least final loss value is shown. In the presence of local material orientation (tows), such material rotations do improve training and validation accuracy, but only marginally. Furthermore, the inclusion of these rotations may increase the \emph{risk} of overfitting and higher test interpolative errors, due to more fitting parameters in $\vtheta_0$ of \eqref{eq:ptTpt}. According to Fig. \ref{fig:woven_training_error}(b), the median and the 0.9-quantile values of the test error are now 4.6\% and 12\%. For the model with the \emph{second} least final loss value, somehow less overfitting is present. The median test error becomes 3\% and its 0.9-quantile is 6.7\%, \emph{i.e.}, still slightly higher than the case when such rotations are not introduced.

\begin{figure}[htbp]
    \centering
    \includegraphics[width=0.95\textwidth]{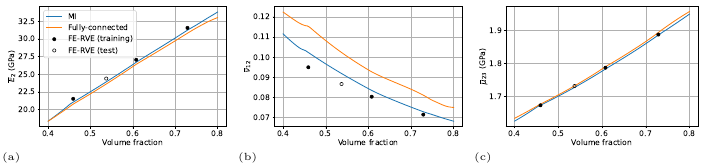}
    \caption{Homogenized elastic moduli prediction as a function of the tow volume fraction obtained by parametric DMN models: (a) in-plane Young's modulus $\overline{E}_2$; (b) in-plane Poisson ratio $\overline{\nu}_{12}$; (c) transverse shear modulus $\overline{\mu}_{23}$.} \label{fig:woven_moduli_pinndmn_fully}
\end{figure}

With the real linear elastic properties provided in Tab. \ref{tab:woven_real_properties}, the homogenized elastic moduli are computed in Fig. \ref{fig:woven_moduli_pinndmn_fully} with varying volume fraction. For comparison, the fully-connected architecture \eqref{eq:fully_connected} is also tested with the physical constraints. Recall that \eqref{eq:fully_connected} implies that the DMN rotations also become a function of the volume fraction. Both models capture well the nonlinear $\vf$-dependence of these elastic moduli. The fully-connected architecture not only increases the number of fitting parameters ($\vec{\Theta}_1$ is now added), it does not improve prediction accuracy compared to the MI one \eqref{eq:physics_informed}. For MIpDMN, the maximum relative Frobenius-norm error $e_i$ in \eqref{eq:loss} is less than 1.5\% among the 4 FE-RVE data points.

\begin{figure}[htbp]
    \centering
    \includegraphics[width=0.8\textwidth]{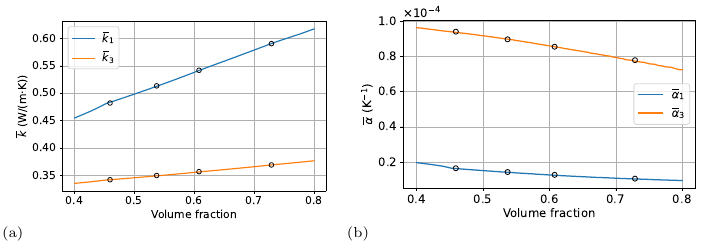}
    \caption{(a) Effective thermal conductivity prediction for the in-plane $\overline{k}_1$ and out-of-plane $\overline{k}_3$ components; (b) effective CTE prediction. The FE-RVE results are indicated by circles.} \label{fig:woven_vf_rev_kBar_alphaBar}
\end{figure}

Using the previously trained 7-layer MIpDMN, we now predict the effective thermal conductivity and the effective CTE in Fig. \ref{fig:woven_vf_rev_kBar_alphaBar} using the real properties of the constituents provided in Tab. \ref{tab:woven_real_properties}. Due to the low CTE values of the tows (fibers) in the woven plane, the effective in-plane $\overline{\alpha}_1$ is well lower than the out-of-plane one. An excellent agreement is found with the FE-RVE results. Among the 4 data points, the maximum relative Frobenius-norm error is 0.25\% for $\kBar$ at $\vf=0.459$ and 0.98\% for $\alphaBar$ also at $\vf=0.459$. Even though MIpDMN is only trained using isothermal linear elastic data, it is capable of predicting other physical properties when the microstructure morphology varies.

\begin{figure}[htbp]
    \centering
    \includegraphics[width=0.4\textwidth]{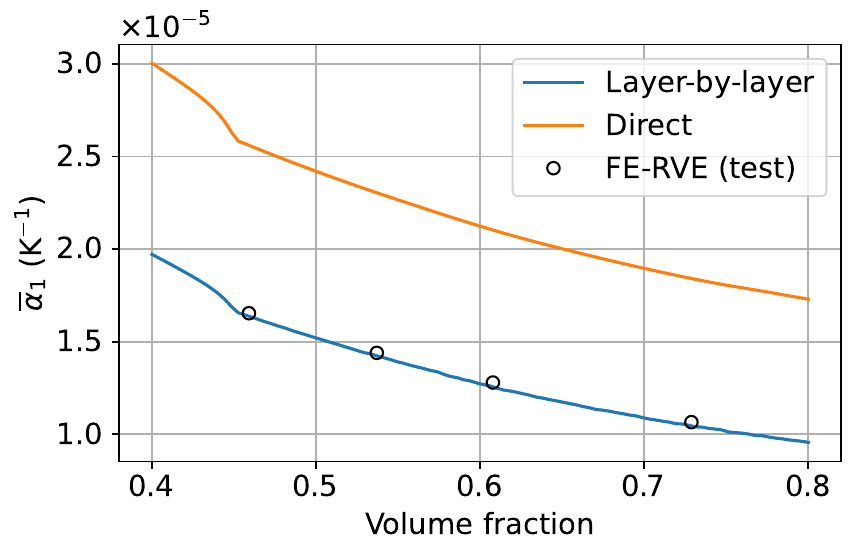}
    \caption{Comparison between a direct application and a DMN layer-by-layer application of the laminate homogenization function of \eqref{eq:laminatealphaBar} for the effective CTE computation.} \label{fig:woven_vf_alphaBar_direct}
\end{figure}

In this case, the second-order CTE tensor of the tows is transversely isotropic in its material frame, while the matrix remains isotropic. This implies that \eqref{eq:laminatealphaBar} must be applied as a neuron operation from the input layer to the output layer along with the effective stiffness tensor computation. If the laminate homogenization function of \eqref{eq:laminatealphaBar} for the effective CTE computation is applied directly to the obtained effective stiffness tensor, as shown in Fig. \ref{fig:woven_vf_alphaBar_direct}, the result would be incorrect.

The computational efficiency of MIpDMN in terms of computing the homogenized stiffness tensor given $\mathbb{C}_1$ and $\mathbb{C}_2$ is compared to FE-RVE simulations in Tab. \ref{tab:speedup}. Thanks to this significant speed-up, MIpDMN can be further used for parametric analysis \cite{Liu:2019b}, uncertainty quantification \cite{Huang:2022} and material property calibration \cite{Dey:2023}. In particular, it can be employed to identify \emph{simultaneously} the \emph{material} and \emph{microstructural} parameters in an inverse identification problem. A preliminary proof-of-concept study is provided in \ref{sec:inverse}.
\begin{table}[htbp]
    \centering
    \begin{tabular}{lll} \toprule
                             & FE-RVE & 7-layer MIpDMN \\ \midrule
        Wall time (speed-up) & 11 s   & 6.62 ms (1662) \\ \bottomrule
    \end{tabular}
    \caption{Computational speed-up of MIpDMN compared to FE-RVE in terms of homogenized stiffness tensor prediction.}
    \label{tab:speedup}
\end{table}

\subsection{Ellipsoidal inclusion composite} \label{sec:ellip}
Finally, we consider the ellipsoidal inclusion composite described by two parameters: volume fraction $\vf$ of the fibers and the aspect ratio $\ar$ (ratio between the length and the diameter of the fiber). The aspect ratio parameter is purely morphological and does not change $\vf$. According to \eqref{eq:pvq}, we have thus $\vec{q}=(\ar)$ in this case.

\begin{figure}[htbp]
    \centering
    \includegraphics[width=0.4\textwidth]{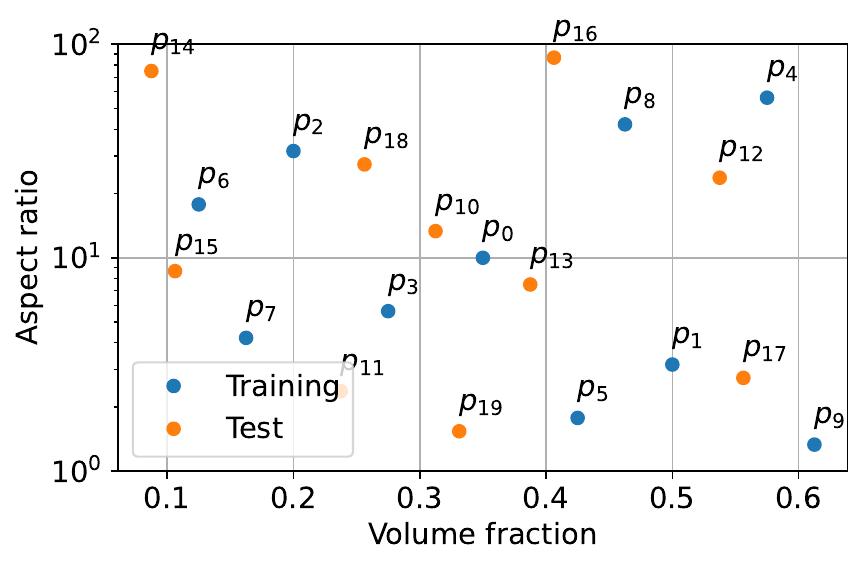}
    \caption{Sampling points in the parametric space for the ellipsoidal inclusion composite.} \label{fig:ellip_doe}
\end{figure}

The two-dimensional parametric space $(\vf,\ar)$ is sampled using Sobol low-discrepancy sequence which ensures an optimal space-filling. Since it is deterministic, additional sampling points can thus be easily included. The bounds for these two parameters are
\begin{itemize}
    \item Volume fraction: $[0.05, 0.065]$. Higher volume fractions would lead to mesh generation issues with the body-centered fiber packing.
    \item Aspect ratio: $[1, 100]$. Both spherical inclusions $\ar=1$ and slender fibers $\ar\gg 1$ are covered with such large variation of this $\ar$ parameter.
\end{itemize}
In this work, 20 samples of $\vec{p}_i=(\vf,\ar)_i$ are generated, see Fig. \ref{fig:ellip_doe}. The $\ar$ parameter is sampled in the log-scale. The first 10 points $\vec{p}_{0\leq i< 10}$ constitute the parametric sample set $\mathbb{P}$ and are used to generate training dataset. The other 10 points $\vec{p}_{10\leq i< 20}$ are reserved to evaluate the generalization accuracy of MIpDMN.

\begin{figure}[htbp]
    \centering
    \includegraphics[width=0.7\textwidth]{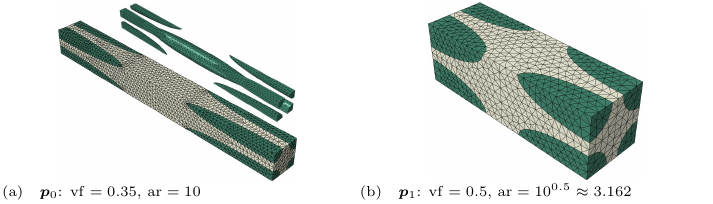}
    \caption{Example FE-RVE models for the ellipsoidal composite with different fiber volume fraction and aspect ratio values: (a) $\vec{p}_0$; (b) $\vec{p}_1$.} \label{fig:ellip_meshes}
\end{figure}

For each $\vec{p}_i=(\vf,\ar)_i$, a 3-d FE-RVE model is built using the body-centered fiber packing. In Fig. \ref{fig:ellip_meshes}, two examples are given corresponding to the first two sampling points. The models contain approximately several hundreds of thousands of quadratic tetrahedral elements and may take up to 60 seconds for each run.

The real linear elastic properties of the two phases are the same as for the unidirectional fiber composite, given in Tab. \ref{tab:hexa_real_properties}. Given the high number of microstructures (10 for training, 10 for test), in total 150 input orthotropic material properties $(\mathbb{C}_1,\mathbb{C}_2)$ are sampled to produce $\mathbb{M}$. The synthetic fiber stiffnesses $\mathbb{C}_2$ are also generated to be statistically higher than the matrix ones $\mathbb{C}_1$. For the training microstructures $\vec{p}_{0\leq i< 10}$, 100 of the 150 samples are used as training dataset, while the other 50 are reserved for validation. For the others $\vec{p}_{10\leq i< 20}$, all the 150 material samples are used for testing the generalization accuracy.

\begin{figure}[htbp]
    \centering
    \includegraphics[width=0.8\textwidth]{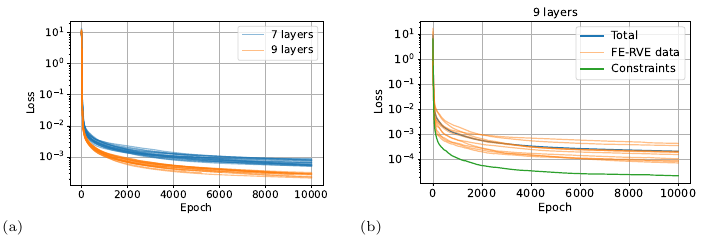}
    \caption{MIpDMN training for the ellipsoidal inclusion composite: (a) loss histories with 7 or 9 DMN layers; (b) partition of the total loss into the FE-RVE data part and the physical constraints part.} \label{fig:ellip_loss}
\end{figure}

MIpDMN is trained following Sect. \ref{sec:training} with the two physical constraints \eqref{eq:pinndmn_vf_loss} and \eqref{eq:pinndmn_aij_loss} and the ReLU activation function. Since local material orientation is absent, the unidirectional material frame orientation tensor \eqref{eq:saij_ud} is defined as the target for both phases in \eqref{eq:pinndmn_aij_loss}. As before, the training process is realized 20 times with random initialization and the model with the least final loss value is chosen for further investigations. The median training time is only 800 s for 7 layers and 1198 s for 9 layers. The loss histories for 7 and 9 DMN layers are presented in Fig. \ref{fig:ellip_loss}. The final loss value is well decreasing with more layers. The FE-RVE data part and the physical constraints part are also monotonically decreasing.

\begin{figure}[htbp]
    \centering
    \includegraphics[width=0.8\textwidth]{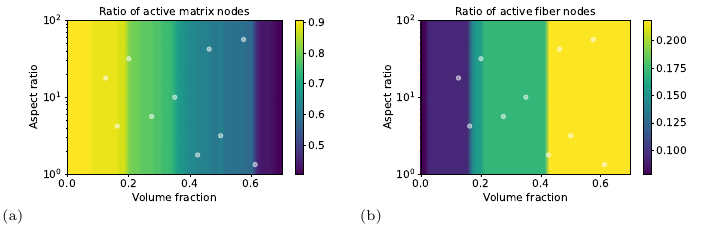}
    \caption{Variation of the ratio of active DMN nodes in the parametric space: (a) matrix; (b) fiber. The training points are indicated by circles.} \label{fig:ellip_n_dof_active}
\end{figure}

The variation of the number of active DMN nodes is presented separately for the matrix and for the fiber in Fig. \ref{fig:ellip_n_dof_active}. 7 DMN layers are used. In total, 73 out of the 128 DMN nodes are globally active in the parametric space $\vec{p}=(\vf,\ar)\in[0, 0.7]\times[1, 100]$. Due to the MI architecture \eqref{eq:physics_informed}, the volume fraction parameter determines the number of active matrix or fiber nodes while the aspect ratio parameter has no influence.

\begin{figure}[htbp]
    \centering
    \includegraphics[width=0.8\textwidth]{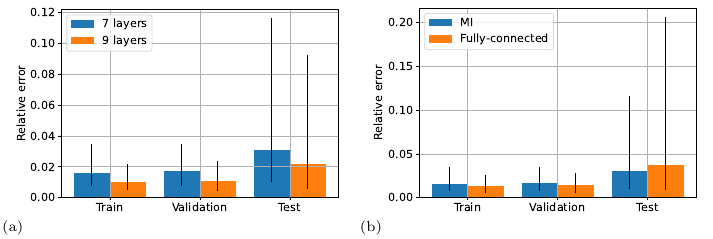}
    \caption{Training, validation and test errors of parametric DMN models: (a) comparison between 7 and 9 layers using the micromechanics-informed (MI) architecture; (b) comparison with the fully-connected architecture using 7 layers.} \label{fig:ellip_training_error}
\end{figure}

In Fig. \ref{fig:ellip_training_error}(a), the training, validation and test errors are presented for MIpDMN with 7 or 9 layers. These errors are first computed at each of the 20 sampling points, then aggregated using quantiles. All these errors become lower with more DMN layers. Overfitting is not observed since the validation errors are comparable with the training ones. For the training and validation errors, their statistical variations are limited. In the example of a 7-layer MIpDMN, the 0.9-quantile error is less than 4\% while the median error value is less than 2\%. However, the test errors present larger statistical dispersion since their 0.9-quantiles reach approximately 10\%. Due to a large amount of test points in the parametric space (see Fig. \ref{fig:ellip_doe}), MIpDMN is being evaluated both with interpolation and extrapolation. The 0.9-quantile of the test errors is mainly due to the presence of ``outliers'' at some particularly challenging test microstructures. The median test errors, nevertheless, are only slightly larger than the training and the validation errors, as before.

These errors are also computed for a fully-connected architecture \eqref{eq:fully_connected} with 7 DMN layers. Compared to the MI one \eqref{eq:physics_informed}, now the DMN fitting parameters become function of both microstructural parameters $\vec{w}=\vec{w}(\vf,\ar)$ and $\vtheta=\vtheta(\vf,\ar)$. The physical constraints \eqref{eq:pinndmn_vf_loss} and \eqref{eq:pinndmn_aij_loss} are included. According to Fig. \ref{fig:ellip_training_error}(b), not only the fully-connected architecture does not significantly reduce training and validation errors, it also nearly doubles the 0.9-quantile of the test errors. The MI architecture \eqref{eq:physics_informed} ensures hence a comparable expressive power in the parametric space with certain generalization capability.

\begin{figure}[htbp]
    \centering
    \includegraphics[width=0.8\textwidth]{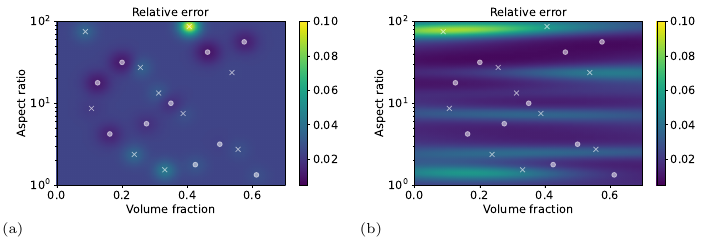}
    \caption{Median errors of parametric DMN models in the parametric space: (a) micromechanics-informed; (b) fully-connected. The training and test points are respectively indicated by circles and crosses.} \label{fig:ellip_training_error_parametric}
\end{figure}

In Fig. \ref{fig:ellip_training_error_parametric}, the median errors are first computed at each of the 20 training and test sampling points, then interpolated and extrapolated to the whole parametric space using Kriging implemented in the SMT library \cite{Bouhlel:2019}. In Fig. \ref{fig:ellip_training_error_parametric}(a), the maximum error (10\%) is localized at the ``outlier'' microstructure $\vec{p}_{16}$ with a very high aspect ratio (87) compared to the training domain. Except this point, the median errors are between 1\% and 4\% and are visually uniform in the parametric space. The fully-connected architecture, however, produces more error variations in the parametric space. This demonstrates again the satisfying interpolation and extrapolation generalization capability of the MI architecture.

\begin{figure}[htbp]
    \centering
    \includegraphics[width=0.95\textwidth]{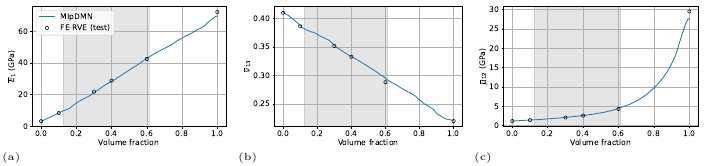}
    \caption{Homogenized elastic moduli prediction as a function of $\vf$ at fixed $\ar=20$ using a 9-layer MIpDMN: (a) longitudinal Young's modulus $\overline{E}_1$; (b) transverse Poisson ratio $\overline{\nu}_{13}$; (c) transverse shear modulus $\overline{\mu}_{12}$. The training domain is represented by a gray shadowed region. The theoretical values at $\vf=0$ and $\vf=1$ are indicated as test data.} \label{fig:ellip_moduli_vf}
\end{figure}

With the real linear elastic properties in Tab. \ref{tab:hexa_real_properties}, the $\vf$-dependence of the homogenized elastic moduli at fixed aspect ratio $\ar=20$ are illustrated in Fig. \ref{fig:ellip_moduli_vf} using a 9-layer MIpDMN. A perfect agreement is found with the additional test FE-RVE simulation data, knowing that they all correspond to previously unseen microstructures during training. The microstructure $\vf=0.1$ lies even outside the training domain. Among the data points, the maximum relative Frobenius-norm error on $\CBar$ is only 3.3\% at $\vf=0.6$.

\begin{figure}[htbp]
    \centering
    \includegraphics[width=0.95\textwidth]{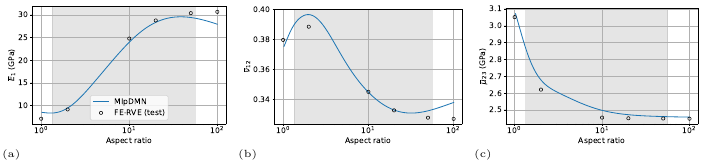}
    \caption{Homogenized elastic moduli prediction as a function of $\ar$ at fixed $\vf=0.4$ using a 9-layer MIpDMN: (a) longitudinal Young's modulus $\overline{E}_1$; (b) transverse Poisson ratio $\overline{\nu}_{12}$; (c) in-plane shear modulus $\overline{\mu}_{23}$. The training domain is represented by a gray shadowed region.} \label{fig:ellip_moduli_ar}
\end{figure}

In Fig. \ref{fig:ellip_moduli_ar}, the influence of the aspect ratio parameter is analyzed with a fixed $\vf=0.4$ and compared with the FE-RVE results on these new microstructures. Not only MIpDMN is capable of predicting the increase of $\overline{E}_1$ with $\ar$ (at least in the training domain), it can also account for the more subtle variations of $\overline{\nu}_{12}$ and $\overline{\mu}_{23}$, which are all in good agreement with the FE-RVE data. Outside the training domain for $\ar$, good predictions are obtained at $\ar=1$, while some deviations are found near $\ar=100$, where the relative Frobenius-norm error is 5.9\%. While the physical constraints can indeed improve interpolative and extrapolative generalization capability of MIpDMN, they cannot replace the input training data. If the prediction at higher aspect ratios is important, more corresponding FE-RVE data could be included during training.

\begin{figure}[htbp]
    \centering
    \includegraphics[width=0.8\textwidth]{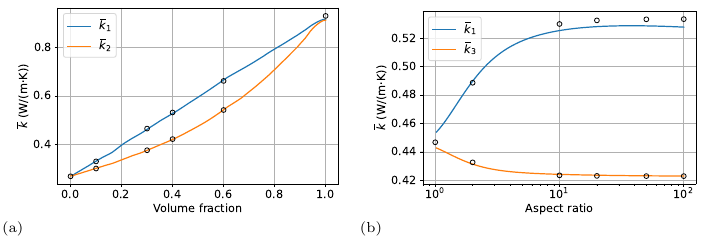}
    \caption{Effective thermal conductivity prediction using a 9-layer MIpDMN for the longitudinal $\overline{k}_1$ and transverse $(\overline{k}_2,\overline{k}_3)$ components: (a) as a function of $\vf$ at fixed $\ar=20$; (b) as a function of $\ar$ at fixed $\vf=0.4$. The FE-RVE results and the theoretical values for $\vf=0$ and $\vf=1$ are indicated by circles.} \label{fig:ellip_vf_ar_rev_kBar}
\end{figure}

Using the real thermal conductivity and CTE properties of the constituents provided previously in Tab. \ref{tab:hexa_real_properties}, we will now use MIpDMN trained with isothermal linear elastic data to predict the effective thermal conductivity and the effective CTE. In Fig. \ref{fig:ellip_vf_ar_rev_kBar}, the nonlinear $\vf$- and $\ar$-dependences of $\kBar$ are respectively well predicted. A satisfying agreement with the FE-RVE results is obtained for both the longitudinal and transverse components. Among the data points, the maximum relative Frobenius-norm error is 0.64\% at $\vf=0.4$ with fixed $\ar=20$, and $1.1\%$ at $\ar=1$ with fixed $\vf=0.4$. This remarkable result illustrates again the multiple physics property prediction capability of DMN, knowing that the thermal conductivity and the CTE data are not used as training data at all.

\begin{figure}[htbp]
    \centering
    \includegraphics[width=0.8\textwidth]{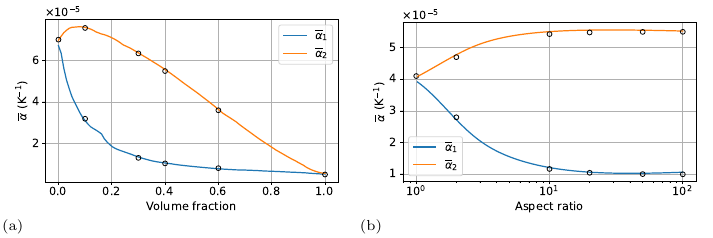}
    \caption{Effective CTE prediction using a 9-layer MIpDMN for the longitudinal $\overline{\alpha}_1$ and transverse $\overline{\alpha}_2$ components: (a) as a function of $\vf$ at fixed $\ar=20$; (b) as a function of $\ar$ at fixed $\vf=0.4$. The FE-RVE results and the theoretical values for $\vf=0$ and $\vf=1$ are indicated by circles.} \label{fig:ellip_vf_ar_rev_alphaBar}
\end{figure}

In Fig. \ref{fig:ellip_vf_ar_rev_alphaBar}, the effective CTE is predicted while varying $\vf$ and $\ar$. As before, a good agreement is found between the MIpDMN prediction and the FE-RVE results. Among the data points, the maximum relative Frobenius-norm error is 2.2\% at $\vf=0.6$ with fixed $\ar=20$, and $4.6\%$ at $\ar=1$ with fixed $\vf=0.4$.

\subsubsection*{Nonlinear mechanical simulations with an elasto-plastic matrix}
\begin{figure}[htbp]
    \centering
    \includegraphics[width=0.4\textwidth]{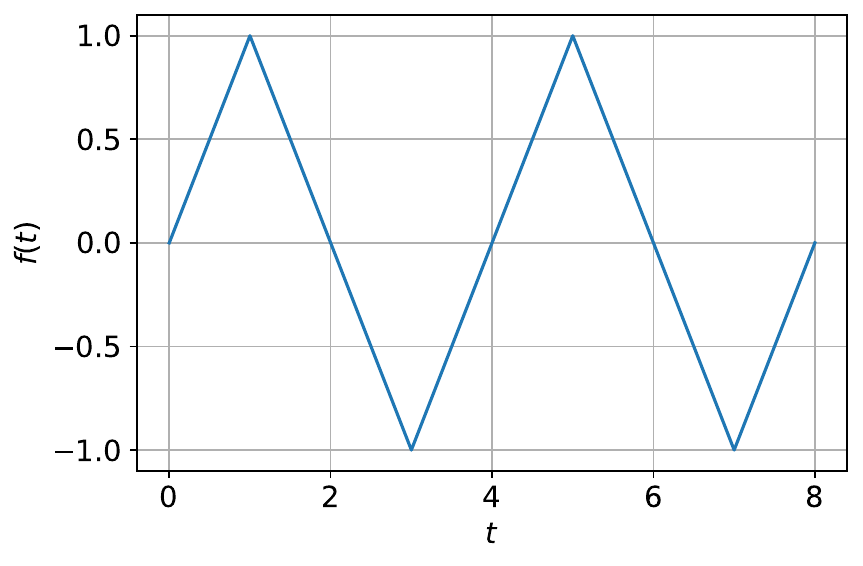}
    \caption{Cyclic function $f(t)$ used in \eqref{eq:multiaxial_cyclic}.} \label{fig:ellip_elastoplastic_rev_loading}
\end{figure}

Using the online prediction formulation described in \ref{sec:nonlinear}, we will now evaluate MIpDMN with nonlinear mechanical material behaviors while varying the microstructural parameters. We assume that the polypropylene matrix follows an elasto-plastic law with an isotropic power-law hardening
\begin{equation*}
    \sigma_\mathrm{Y}=\sigma_0+k(p+\epsilon)^n,
\end{equation*}
where $\sigma_\mathrm{Y}$ defines the yield surface, $\sigma_0$ is the yield stress, and $(k,n)$ describes strain hardening with the equivalent plastic strain $p$. The following numerical values are used: $\sigma_0=30$ MPa, $k=293$ MPa and $n=0.34$. The small value $\epsilon=10^{-6}$ is used to avoid infinite derivative $\mathrm{d}\sigma_\mathrm{Y}/\mathrm{d}p$ at $p=0$. The following proportional cyclic multiaxial loading--unloading path is considered in the strain space
\begin{equation} \label{eq:multiaxial_cyclic}
    \epsBar(t)=f(t)\epsBar_0,\quad \epsBar_0=(\overline{\varepsilon}_{11},\overline{\varepsilon}_{22},\overline{\varepsilon}_{33},\sqrt{2}\overline{\varepsilon}_{12},0,0),\quad \overline{\varepsilon}_{11}=1\%,\,\overline{\varepsilon}_{22}=\overline{\varepsilon}_{12}=4\%,\,\overline{\varepsilon}_{33}=-2\%.
\end{equation}
The cyclic function $f(t)$ defined on $[0,T]$ is presented in Fig. \ref{fig:ellip_elastoplastic_rev_loading}. The time parameter $t$ is assumed dimensionless since quasi-static loading is considered.

\begin{figure}[htbp]
    \centering
    \includegraphics[width=0.95\textwidth]{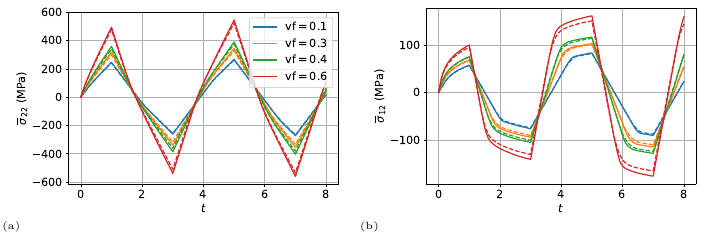}
    \caption{Effective nonlinear behavior of the ellipsoidal inclusion composite for different fiber volume fractions at fixed $\ar=20$ using a 9-layer MIpDMN: (a) $\overline{\sigma}_{22}$; (b) $\overline{\sigma}_{12}$. The DMN predictions are represented by solid lines, while the FE-RVE data are dashed lines.} \label{fig:ellip_elastiplastic_rev_vf}
\end{figure}

In Fig. \ref{fig:ellip_elastiplastic_rev_vf}, the corresponding temporal stress responses are presented for different volume fraction values. The influence of the $\vf$-parameter on the nonlinear effective responses is well captured by MIpDMN. With increasing volume fraction, the maximum value of $\overline{\sigma}_{22}$ increases from approximately 250 MPa to nearly 600 MPa. Furthermore, more plasticity effects are visible especially on the $\overline{\sigma}_{12}$ components. Good agreement between MIpDMN and the FE-RVE results is observed, even though these microstructures are previously unseen during offline training. The difference between them can be quantified with the following error measure on the full stress tensor with the Frobenius norm
\begin{equation} \label{eq:errortemporal}
    e=\frac{\int_0^T\norm{y^\mathrm{DMN}(t)-y^\mathrm{FE}(t)}\,\mathrm{d}t}{\int_0^T\norm{y^\mathrm{FE}(t)}\,\mathrm{d}t}.
\end{equation}
For $\vf=0.1$ and $\vf=0.6$, the errors are only respectively 2.2\% and 5.6\%, considering that these two microstructures lie on the boundary of or even outside the parametric training domain.

\begin{figure}[htbp]
    \centering
    \includegraphics[width=0.95\textwidth]{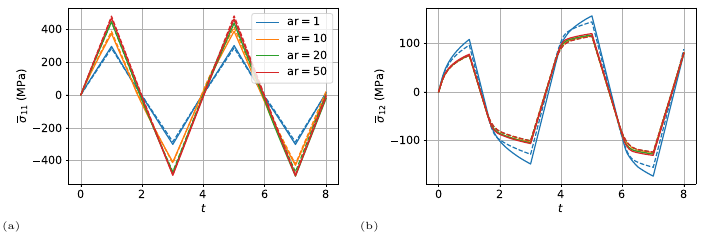}
    \caption{Effective nonlinear behavior of the ellipsoidal inclusion composite for different fiber aspect ratios at fixed $\vf=0.4$ using a 9-layer MIpDMN: (a) $\overline{\sigma}_{11}$; (b) $\overline{\sigma}_{12}$. The DMN predictions are represented by solid lines, while the FE-RVE data are dashed lines.} \label{fig:ellip_elastiplastic_rev_ar}
\end{figure}

The $\ar$-dependence of the nonlinear effective behaviors is then analyzed in Fig. \ref{fig:ellip_elastiplastic_rev_ar}. The aspect ratio parameter has a bigger influence on the longitudinal effective stress, which increases from 300 MPa to nearly 500 MPa when $\ar$ varies from 1 (spherical inclusions) to 50 (slender fibers). The transverse shear stress $\overline{\sigma}_{12}$ seems to converge within the given $\ar$ range. Again, MIpDMN is in good agreement with the FE-RVE responses, even though a visible difference can be noticed on $\overline{\sigma}_{12}$ for $\ar=1$. For this microstructure outside the parametric training domain, the error is 8.6\% using \eqref{eq:errortemporal}.

\begin{figure}[htbp]
    \centering
    \includegraphics[width=0.95\textwidth]{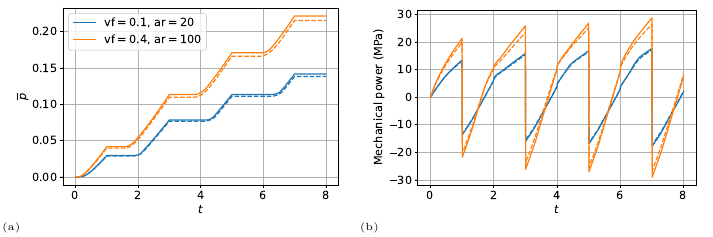}
    \caption{(a) Volume-averaged equivalent plastic strain; (b) Mechanical power. The DMN predictions are represented by solid lines, while the FE-RVE data are dashed lines.} \label{fig:ellip_elastoplastic_rev_peeq_power}
\end{figure}

To further demonstrate the nonlinear prediction capability of MIpDMN, the volume-averaged equivalent plastic strain is computed and compared with the FE-RVE results in Fig. \ref{fig:ellip_elastoplastic_rev_peeq_power}(a). Two microstructures outside the parametric training domain, \emph{i.e.} extrapolating $(\vf,\ar)$, are considered. For DMN, volume-averaging is performed on $p_i$ stored on the material nodes using their weights. Using \eqref{eq:errortemporal} with the absolute-value norm, the errors are only 2.3\% for $\vf=0.1$, $\ar=20$ and 3.1\% for $\mathrm{vf}=0.4$, $\mathrm{ar}=100$.

Inspired by the work of \cite{Nguyen:2022a}, we investigate the energy balance and consistency of MIpDMN during nonlinear online prediction. The (average) mechanical power inside an increment $[t_n,t_{n+1}]$ of duration $\Delta t$ can be computed by
\begin{equation} \label{eq:mechanical_power}
    \mathcal{P}=\frac{1}{\Delta t}\int_{t_n}^{t_{n+1}}\sigBar\cdot\dot{\epsBar}\,\mathrm{d}t=\frac{1}{2\Delta t}\left((\sigBar_{n}+\sigBar_{n+1})\cdot\Delta\epsBar\right).
\end{equation}
In Fig. \ref{fig:ellip_elastoplastic_rev_peeq_power}(b), The MIpDMN mechanical power evolution is compared with the FE-RVE one. A relatively good agreement is also found for these two extrapolating microstructures, since \eqref{eq:errortemporal} give respectively 2.8\% and 8.7\%. For DMN, the effective mechanical power \eqref{eq:mechanical_power} can also be retrieved using the weighted-average of $\mathcal{P}_i$ on the material nodes with their local stress and strain states \cite{Nguyen:2022a}. The errors between this material nodes-based approach and \eqref{eq:mechanical_power} are only 0.70\% for $\vf=0.1$, $\ar=20$ and 1.1\% for $\mathrm{vf}=0.4$, $\mathrm{ar}=100$. This demonstrates the energetic consistency of MIpDMN even when extrapolating the microstructural parameters.

Finally, the computational efficiency of MIpDMN is presented in Tab. \ref{tab:speedup_nonlinear} for the previous two microstructures. Even with a 9-layer architecture, a speed-up of approximately 700 is obtained compared to FE-RVE. Similar factors have also been reported for instance in \cite{Liu:2019a,Gajek:2020,Nguyen:2022a,Dey:2022}. We also recall that the generation of the costly training data on new microstructures even outside the parametric training domain is also no longer needed with MIpDMN.
\begin{table}[htbp]
    \centering
    \begin{tabular}{lll} \toprule
        Wall time (speed-up) & FE-RVE & 9-layer MIpDMN \\ \midrule
        $\vf=0.1$, $\ar=20$  & 3083 s & 4.10 s (752)   \\
        $\vf=0.4$, $\ar=100$ & 3723 s & 5.69 s (654)   \\ \bottomrule
    \end{tabular}
    \caption{Computational speed-up of MIpDMN compared to FE-RVE for the two extrapolating microstructures during the nonlinear online prediction.}
    \label{tab:speedup_nonlinear}
\end{table}

\section{Conclusions} \label{sec:conclusion}
A novel micromechanics-informed parametric deep material network (MIpDMN) architecture is proposed for heterogeneous materials with a varying morphology described by multiple parameters. The dependence of the DMN fitting parameters on the microstructural ones is accounted for by a single layer feedforward neural network informed by micromechanical properties \eqref{eq:physics_informed}. While the DMN weights only vary with the scalar volume fraction parameter \eqref{eq:pwWpz}, the DMN rotations are influenced by other purely morphological parameters that do not change the volume fraction \eqref{eq:ptTpt}. Furthermore, two additional micromechanical constraints \eqref{eq:pinndmn_vf_loss} and \eqref{eq:pinndmn_aij_loss} are prescribed so that micromechanics-informed neural network recovers the volume fraction and the material frame orientation tensors of the actual parameterized microstructure. A unique offline training of MIpDMN is required, based on a total loss function \eqref{eq:loss} that aggregates the FE-RVE data at some given microstructural parameter values. The trained MIpDMN is then able to predict the material behaviors in the parametric space, \emph{i.e.}, even for previously unseen microstructures during training. The time-consuming generation of the corresponding new computational homogenization data is hence no longer needed.

Based on the numerical results reported, the micromechanics-informed architecture as well as the constraints improve generalization capabilities in the parametric space while successfully capturing the structure-property relationships.
The linear and nonlinear effective behaviors of the \emph{parameterized} microstructure in question can hence be accurately predicted by MIpDMN. Besides, we show that the inclusion of initial material rotation matrices on the input layer does not significantly increase the expressive power even for microstructures with local material orientation.

MIpDMN is also recast in a multiple physics setting. Through the redefinition of the laminate homogenization function, other physical properties such as the thermal conductivity and the coefficient of thermal expansion can be accurately predicted at the online prediction stage. Based on the numerical simulations reported, the real relative error on $\alphaBar$ is of the same order of magnitude as $\CBar$ (approximately 1\% to 6\%), while that on $\kBar$ is in general near or less than 1\%. This is a remarkable result since MIpDMN is trained using isothermal linear elastic data. It demonstrates that our MIpDMN learns the \emph{parameterized} microstructure \emph{per se}, and not a physical property in particular.

The limitations of this work as well as some possible future directions are summarized as follows:
\begin{itemize}
    \item Even though \eqref{eq:physics_informed} extends naturally to higher parametric dimensions, at most only two microstructural parameters are considered in the numerical simulations. The microstructures with more morphological parameters will be investigated in the future to demonstrate its full potential with respect to the existing approaches.
    \item A single-layer architecture is considered in \eqref{eq:physics_informed}. For more complex structure-property relationships, it can be expected that $\vec{q}\mapsto\vtheta(\vec{q})$ in \eqref{eq:ptTpt} may become nonlinear and require hidden layers and weight regularization. It will be investigated in our future work.
    \item The MIpDMN formulation presented in this paper mainly targets two-phase parameterized microstructures. For polycrystalline materials, the partition \eqref{eq:pvq} needs to be redefined knowing that now the 1-point correlation function becomes the crystallite orientation distribution function (CODF), see \cite{Boehlke:2005}. Hence, it suggests that in the MI architecture \eqref{eq:physics_informed}, now the DMN weights would become a function of CODF. Interestingly, it has been numerically found in \cite{Liu:2019a} that the $\vec{w}$ indeed remain uniformly distributed for a uniform (totally random) CODF and vary with the texture information.
    \item For multiphase materials, the non-parametric DMN architecture can be first generalized by considering multiphase laminates as building blocks \cite{Liu:2019a,Gajek:2020} or an interaction-based network \cite{Nguyen:2022}. Then, the partition \eqref{eq:pvq} can be adapted by using correlation functions for multiphase materials. For instance, the single $\vf$ parameter would be replaced by $P-1$ volume fractions for a $P$-phase microstructure. Also, the physical constraints \eqref{eq:pinndmn_vf_loss} and \eqref{eq:pinndmn_aij_loss} need to be accordingly extended for the additional constituents.
    \item In this work, we focused on the extension of the original DMN formulation to parameterized microstructures. Hence, much more complex constitutive behaviors such as viscoplasticity \cite{Dey:2022} are not considered. Furthermore, accounting for localized material failure such as fracture is also challenging especially for DMN. Compared to other reduced-order models such as the self-consistent clustering analysis \cite{Liu:2018}, spatial fields in the RVE are volume-averaged to the DMN material nodes. In \cite{Liu:2021}, a cell division scheme is proposed for consistent scale transition in DMN with promising results. It would be interesting in the future to couple it with non-local damage models such as phase-field fracture \cite{Bharali:2021} to tackle multiscale strain-localization problems.
    \item We investigated the use of MIpDMN trained using stiffness tensors $\tens{C}$ to predict other physical properties such as thermal conductivity $\vec{k}$. Our preliminary results not reported here indicate that MIpDMN trained with $\vec{k}$ can also predict $\tens{C}$, with a slightly decreased accuracy. Future work will be devoted to the offline training of MIpDMN using \emph{combined} multiple physics data.
\end{itemize}

\section*{Acknowledgments}
The author thanks Rafael SALAZAR TIO for the fruitful discussions on using DMN in a multiple physics setting. During her internship, Maëlle PRAUD helped with investigating various optimization algorithms.

\appendix

\section{Nonlinear homogenization formulation for online prediction} \label{sec:nonlinear}
Consider a time interval $[t_n,t_{n+1}]$ and an increment in the macroscopic strain from $\epsBar_n$ to $\epsBar_{n+1}=\epsBar_n+\Delta\epsBar$. As explained in Sect. \ref{sec:network}, a material node is attached to each phase of the leaf laminates, see Fig. \ref{fig:dmn_laminate}. At time $t_n$, we suppose that all their material state variables (strains, stresses and internal state variables) are known. The objective of the DMN surrogate during nonlinear online prediction is to compute at $t_{n+1}$ the new homogenized stress tensor and the consistent tangent operator
\begin{equation} \label{eq:nonlinearEq}
    \Delta\epsBar\mapsto(\CBar,\sigBar_{n+1}).
\end{equation}

As for mean-field homogenization approaches, a linearization scheme is required \cite{Kanoute:2009} to linearize nonlinear constitutive laws. In \cite{Liu:2019a}, an incrementally affine formulation similar to \cite{Doghri:2010} is proposed for DMN. In this work, we continue to use this nonlinear homogenization formulation, since it leads to a fixed-point problem that can be efficiently solved by acceleration methods \cite{Ramiere:2015}.

The incrementally affine formulation assumes that for each material node, the stress increment in this time interval can be computed as follows
\begin{equation} \label{eq:Dsig}
    \sig_{n+1}-\sig_n=\Delta\sig=\tens{C}\Delta\eps+\delta\sig,
\end{equation}
where $\Delta\eps$ is the strain increment for this material node, $\delta\sig$ the residual stress increment which measures material nonlinearity in this time interval and $\tens{C}$ is the consistent tangent operator. Note that in \eqref{eq:Dsig}, the strain increment $\Delta\eps$ is also unknown, since only the increment in the macroscopic strain tensor $\Delta\epsBar$ is known. A consequence of \eqref{eq:Dsig} is that the effective stress tensor increment can also be put in this incrementally affine form
\begin{equation} \label{eq:dmn_sigBar}
    \sigBar_{n+1}-\sigBar_n=\Delta\sigBar=\CBar\Delta\epsBar+\overline{\delta\sig}.
\end{equation}
The computation of \eqref{eq:nonlinearEq} involves the following three \emph{coupled} operations in the DMN architecture, see \cite{Liu:2019a}.
\begin{enumerate}
    \item Forward homogenization: The consistent tangent operators $\tens{C}$ and residual stresses $\delta\sig$ defined on each material node are homogenized from the leaf laminates to the root laminate, obtaining the homogenized tensors $\CBar_l$ and $\overline{\delta\sig}_l$ for each laminate
          \begin{equation} \label{eq:forward}
              \operatorname{Forward}:(\tens{C}_i,\delta\sig_i)_{1\leq i\leq\nnodes}\mapsto(\CBar_l,\overline{\delta\sig}_l).
          \end{equation}
    \item Backward localization: The strain increment for each material node $\Delta\eps_i$ is computed based on the macroscopic strain increment $\Delta\epsBar$ and the strain concentration tensors defined in each laminate obtained during forward homogenization, from the root laminate to the leaf laminates
          \begin{equation} \label{eq:backward}
              \operatorname{Backward}:(\CBar_l,\overline{\delta\sig}_l;\Delta\epsBar)\mapsto(\Delta\eps_i)_{1\leq i\leq\nnodes}.
          \end{equation}
    \item Constitutive behavior integration: For each material node, $\tens{C}_i$ and $\delta\sig_i$ are computed by integrating constitutive behaviors, given their respective strain increment $\Delta\eps_i$.
          \begin{equation} \label{eq:matint}
              \operatorname{MatInt}:(\Delta\eps_i)_{1\leq i\leq\nnodes}\mapsto(\tens{C}_i,\delta\sig_i)_{1\leq i\leq\nnodes}.
          \end{equation}
\end{enumerate}

These coupled equations \eqref{eq:forward}, \eqref{eq:backward} and \eqref{eq:matint} naturally give rise to a \emph{vectorial} fixed point problem $\underline{x}=\mathsf{F}(\underline{x})$ on the flattened vector of strain increments $\underline{x}=(\Delta\eps_i)_{1\leq i\leq\nnodes}$ for the time interval $[t_n,t_{n+1}]$
\begin{equation} \label{eq:fixed_point}
    \underline{x}=(\underbrace{\operatorname{Backward}\circ\operatorname{Forward}\circ\operatorname{MatInt}}_{\mathsf{F}})(\underline{x})\implies \underline{x}^{j+1}=\mathsf{F}(\underline{x}^j)
\end{equation}
where $j\geq 0$ is the fixed-point iteration number.

In this work, acceleration methods \cite{Ramiere:2015} are used to accelerate the convergence of the fixed-point problem \eqref{eq:fixed_point}. In particular, Aitken relaxation \cite{Irons:1969} is applied as a post-processing step in \eqref{eq:fixed_point}
\begin{equation} \label{eq:aitken}
    \begin{aligned}
        \underline{x}^{j+1} & = \mathsf{F}(\underline{x}^j)+(\omega^j-1)\underline{r}^j,\quad \underline{r}^j=\mathsf{F}(\underline{x}^j)-\underline{x}^j                    \\
        \omega^j            & = \begin{cases}
                                    \omega^0                                                                                                                           & j=0     \\
                                    -\omega^{j-1}\dfrac{(\underline{r}^{j-1})\cdot(\underline{r}^j-\underline{r}^{j-1})}{\norm{\underline{r}^j-\underline{r}^{j-1}}^2} & j\geq 1
                                \end{cases}
    \end{aligned}
\end{equation}
In \eqref{eq:aitken}, after updating the relaxation parameter $\omega^j$, it is then numerically bounded by $[\omega_\mathrm{min},\omega_\mathrm{max}]$. The algorithm that realizes nonlinear homogenization \eqref{eq:nonlinearEq} in $[t_n,t_{n+1}]$ is summarized as follows.
\begin{itemize}
    \item Given macroscopic strain increment $\Delta\epsBar$ with $\norm{\Delta\epsBar}>\epsilon$. In this work we use $\epsilon=10^{-8}$.
    \item Initialize the strain increments $\underline{x}^0$ on material nodes, by applying backward localization \eqref{eq:backward} with the strain concentration tensors computed at the last time increment $t_n$. For the first time increment $t_0$, each material node and laminate is initialized using linear elastic properties.
    \item For the fixed-point iteration $j\geq 0$:
          \begin{enumerate}
              \item Integrate constitutive behaviors for each material node \eqref{eq:matint}, obtaining $(\tens{C}_i,\delta\sig_i)_{1\leq i\leq\nnodes}$.
              \item Forward homogenization \eqref{eq:forward}, obtaining $(\CBar_l,\overline{\delta\sig}_l)$.
              \item Backward localization \eqref{eq:backward}, obtaining $\mathsf{F}(\underline{x}^j)$.
              \item Check convergence. In this work, the relative error in the $L_2$ norm with respect to the macroscopic strain increment is controlled
                    \begin{equation} \label{eq:erel}
                        e_\mathrm{rel}=\frac{\norm{\mathsf{F}(\underline{x}^j)-\underline{x}^j}}{\norm{\Delta\epsBar}}<\mathrm{rtol}.
                    \end{equation}
                    If convergence is reached, end the fixed-point iterations.
              \item Apply Aitken relaxation \eqref{eq:aitken}, obtaining $\underline{x}^{j+1}$.
          \end{enumerate}
    \item Compute the effective stress tensor \eqref{eq:dmn_sigBar}.
\end{itemize}
If all the material nodes remain linear elastic, only 1 fixed-point iteration is required, since $\norm{\mathsf{F}(\underline{x}^0)-\underline{x}^0}=0$ in \eqref{eq:erel}.

\begin{figure}[htbp]
    \centering
    \includegraphics[width=0.4\textwidth]{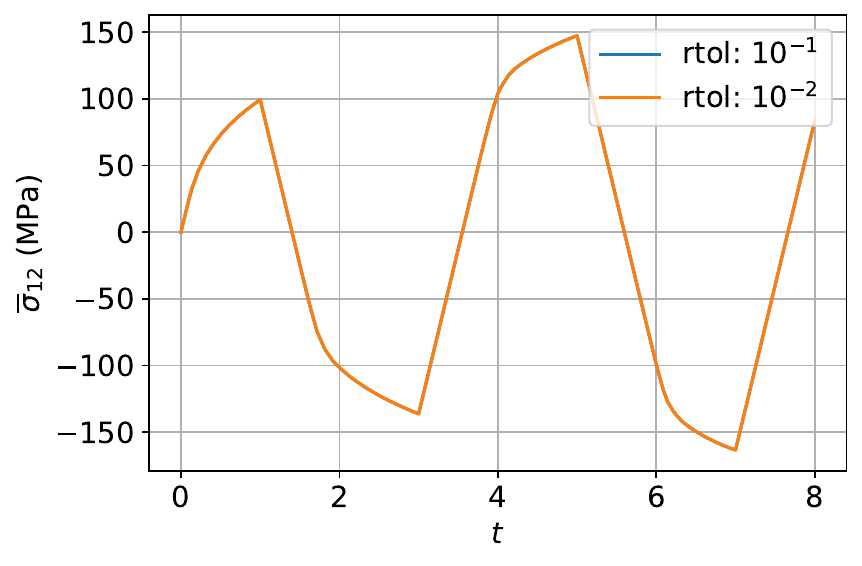}
    \caption{Effect of the tolerance value on the obtained nonlinear effective behavior for the ellipsoidal inclusion composite.} \label{fig:ellip_elastoplastic_rtol}
\end{figure}

We propose to numerically investigate the use of Aitken relaxation \eqref{eq:aitken} on the convergence of the fixed-point problem \eqref{eq:fixed_point}, by using the cyclic multiaxial loading problem \eqref{eq:multiaxial_cyclic} on the ellipsoidal inclusion composite presented in Sect. \ref{sec:ellip}. First, two tolerance values in \eqref{eq:erel} are compared in Fig. \ref{fig:ellip_elastoplastic_rtol}, with $\vf=0.4$ and $\ar=2$. We observe that $\mathrm{rtol}=10^{-1}$ is sufficient to obtain converged nonlinear responses.

\begin{figure}[htbp]
    \centering
    \includegraphics[width=0.8\textwidth]{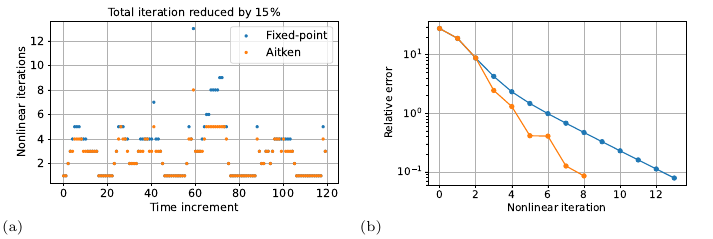}
    \caption{Influence of Aitken relaxation on the convergence of the nonlinear fixed-point problem: (a) number of fixed-point iterations for each time increment; (b) relative error history for the increment 59.} \label{fig:ellip_elastoplastic_aitken}
\end{figure}

In Fig. \ref{fig:ellip_elastoplastic_aitken}(a), the number of fixed-point iterations is presented for each time increment. In \eqref{eq:aitken}, we use $\omega_0=1$, $\omega_\mathrm{min}=1$ and $\omega_\mathrm{max}=2$. This means we are over-relaxing the already converging fixed-point iterations. When the material behavior is linear, only one iteration is required. Otherwise, we observe an overall reduction in the number of iterations with Aitken relaxation. The total number of iterations is reduced by 15\%, which leads to a reduction in the computational time by approximately 13\%. Since Aitken relaxation essentially only involves a scalar product and a norm computation, it is hence much less costly than each fixed-point iteration.

The history of the relative error \eqref{eq:erel} at time increment 59 is used to illustrate the effect of Aitken relaxation, in Fig. \ref{fig:ellip_elastoplastic_aitken}(b). Aitken relaxation accelerates convergence and the number of fixed-point iterations is reduced from 13 to 8.

\section{Inverse identification of unknown material and microstructural parameters} \label{sec:inverse}
Trained MIpDMN can serve as an accurate and computationally efficient surrogate of the parameterized microstructure. Not only it can be used in forward prediction, it can also be employed to identify \emph{simultaneously} the \emph{material} and \emph{microstructural} properties in an inverse identification problem. Below, we provide a preliminary proof-of-concept study for the woven composite presented in Sect. \ref{sec:woven}. Since the objective here is to show that \emph{microstructural} parameters can also be identified, only linear elastic behaviors are considered. For a more realistic inverse identification problem with complex nonlinear material responses, readers can refer to \cite{Dey:2023}.

Now, the effective stiffness tensor $\CBar$ is provided, and the elastic properties of the matrix $\mathbb{C}_1$ and the tows $\mathbb{C}_2$ as well as the volume fraction of the tows $\vf$ are to be identified. In practice $\CBar$ can be measured experimentally. Here we use the FE-RVE simulation result $\CBar^\mathrm{Data}$ on the test microstructure $\vf=0.537$, obtained with the real properties in Tab. \ref{tab:woven_real_properties} which are now sought for.

Motivated by the fact that the gradients with respect to $(\mathbb{C}_1,\mathbb{C}_2,\vf)$ can be easily computed using automatic differentiation thanks to the MIpDMN architecture, in this work we adopt a gradient-based optimization approach based on a loss function \eqref{eq:loss_calibration} which compares DMN prediction with $\CBar^\mathrm{Data}$ using the Frobenius norm. Similar to the training strategy of MIpDMN, the Rprop optimizer is used for minimization.
\begin{equation} \label{eq:loss_calibration}
    \mathcal{L}_\mathrm{cal}=\frac{\norm{\CBar^\mathrm{DMN}-\CBar^\mathrm{Data}}^2}{\norm{\CBar^\mathrm{Data}}^2}.
\end{equation}

The optimization iterations require the initial guess values for the unknowns $(\mathbb{C}_1,\mathbb{C}_2,\vf)$. Hence, the real properties in Tab. \ref{tab:woven_real_properties} as well as the true volume fraction $\vf=0.537$ are randomly perturbed using a normal distribution with a coefficient of variation equal to 20\%. The initial values generated by two random realizations are indicated in Tab. \ref{tab:woven_calibration_init}.
\begin{table}[htbp]
    \centering
    \begin{tabular}{lll} \toprule
        Matrix        & $E$ (MPa) & $\nu$ \\ \midrule
        Realization 1 & 4491      & 0.400 \\
        Realization 2 & 3908      & 0.437 \\ \bottomrule
    \end{tabular} \\ \vspace{0.2cm}
    \begin{tabular}{lllllll} \toprule
        Tow           & $\vf$ & $E_1$ (GPa) & $E_2$ (GPa) & $\nu_{12}$ & $\nu_{23}$ & $\mu_{12}$ (GPa) \\ \midrule
        Realization 1 & 0.550 & 92.3        & 5.60        & 0.347      & 0.697      & 2.14             \\
        Realization 2 & 0.541 & 65.4        & 5.22        & 0.297      & 0.764      & 1.78             \\ \bottomrule
    \end{tabular}
    \caption{Initial linear elastic properties of the two phases for the inverse identification problem.} \label{tab:woven_calibration_init}
\end{table}

\begin{figure}[htbp]
    \centering
    \includegraphics[width=0.8\textwidth]{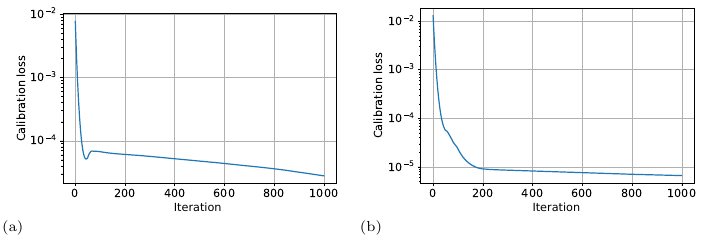}
    \caption{Calibration loss histories using two initial guess values: (a) realization 1; (b) realization 2.} \label{fig:woven_calib_loss}
\end{figure}

The loss histories corresponding to these two initial guess values are given in Fig. \ref{fig:woven_calib_loss}. The loss function converges quickly and may reach $10^{-5}$ within 1000 optimization iterations (a few seconds). The effective stiffness tensor $\CBar$ is well recovered. According to Tab. \ref{tab:woven_calibration_results}, the relative error between the converged effective stiffness tensor and the $\CBar$ data is 0.42\% and 0.26\% for these two sets of initial guess values.
\begin{table}[htbp]
    \centering
    \begin{tabular}{llllllll} \toprule
        $\CBar$       & $E_1=E_2$ (GPa) & $E_3$ (GPa) & $\nu_{12}$ & $\nu_{13}=\nu_{23}$ & $\mu_{12}$ (GPa) & $\mu_{13}=\mu_{23}$ (GPa) & Error  \\ \midrule
        Data          & 24.4            & 7.03        & 0.0866     & 0.558               & 1.87             & 1.73                      &        \\
        Realization 1 & 24.5, 24.3      & 7.05        & 0.0879     & 0.556, 0.555        & 1.87             & 1.73                      & 0.42\% \\
        Realization 2 & 24.4            & 7.02        & 0.0863     & 0.560, 0.558        & 1.86             & 1.74                      & 0.26\% \\ \bottomrule
    \end{tabular}
    \caption{Converged homogenized stiffness tensor from two different realizations of initial values, compared with the true data.} \label{tab:woven_calibration_results}
\end{table}

In Tab. \ref{tab:woven_calibration_results_prop}, the inversely identified input parameters $(\mathbb{C}_1,\mathbb{C}_2,\vf)$ are reported. The material and microstructural parameters found with the realization 1 are similar to the actual properties in Tab. \ref{tab:woven_real_properties}. However, with the realization 2, the identified volume fraction 0.667 is higher than the data 0.537. Meanwhile, the longitudinal Young's modulus (64.6 GPa) of the tows is also smaller than the data (78.8 GPa). This illustrates the non-uniqueness of the inverse identification problem. Additional conditions or data (for instance, nonlinear responses) could be provided to further constrain the inverse problem.
\begin{table}[htbp]
    \centering
    \begin{tabular}{lll} \toprule
        Matrix        & $E$ (MPa) & $\nu$ \\ \midrule
        Realization 1 & 4309      & 0.385 \\
        Realization 2 & 4624      & 0.375 \\ \bottomrule
    \end{tabular} \\ \vspace{0.2cm}
    \begin{tabular}{lllllll} \toprule
        Tow           & $\vf$ & $E_1$ (GPa) & $E_2$ (GPa) & $\nu_{12}$ & $\nu_{23}$ & $\mu_{12}$ (GPa) \\ \midrule
        Realization 1 & 0.533 & 80.7        & 5.53        & 0.340      & 0.614      & 2.09             \\
        Realization 2 & 0.667 & 64.6        & 5.27        & 0.364      & 0.617      & 1.90             \\ \bottomrule
    \end{tabular}
    \caption{Identified linear elastic properties of the two phases and the volume fraction for the inverse identification problem.} \label{tab:woven_calibration_results_prop}
\end{table}

\bibliographystyle{elsarticle-num}
\bibliography{library}
\end{document}